\documentclass[sigconf]{acmart}
\AtBeginDocument{%
  }

\usepackage{hyperref}
\usepackage{url}
\usepackage{multirow}
\usepackage{booktabs}
\usepackage{threeparttable}
\usepackage{graphicx}
\usepackage{xspace}
\usepackage{algorithm}
\usepackage{algorithmic}
\usepackage{xcolor,colortbl}
\usepackage{todonotes}
\usepackage{paralist}
\usepackage{tikz}
\usepackage{etoolbox} 
\usepackage{natbib}

\makeatletter
\newcommand{\newinclude}[1]{%
    \begingroup
    \let\clearpage\relax 
    \input{#1}%
    \endgroup
}
\makeatother

\usepackage{amsmath,amsfonts,bm}

\def\eqref#1{equation~\ref{#1}}
\def\Eqref#1{Equation~\ref{#1}}

\def\1{\bm{1}}

\def\vtheta{{\bm{\theta}}}

\def\vm{{\bm{m}}}

\def\vx{{\bm{x}}}

\DeclareMathAlphabet{\mathsfit}{\encodingdefault}{\sfdefault}{m}{sl}
\SetMathAlphabet{\mathsfit}{bold}{\encodingdefault}{\sfdefault}{bx}{n}

\newcommand{\R}{\mathbb{R}}

\copyrightyear{2025}
\acmYear{2025}
\setcopyright{cc}
\setcctype{by}
\acmConference[CCS '25]{Proceedings of the 2025 ACM SIGSAC Conference on Computer and Communications Security}{October 13--17, 2025}{Taipei, Taiwan}
\acmBooktitle{Proceedings of the 2025 ACM SIGSAC Conference on Computer and Communications Security (CCS '25), October 13--17, 2025, Taipei, Taiwan}\acmDOI{10.1145/3719027.3744846}
\acmISBN{979-8-4007-1525-9/2025/10}

\newcommand{\pie}[1]{%
\begin{tikzpicture}
 \draw (0,0) circle (0.75ex);\fill (0.75ex,0) arc (0:(-#1+90):0.75ex) -- (0,0) -- cycle;
 \fill (0.75ex,0) arc (0:(#1-90):0.75ex) -- (0,0) -- cycle;
\end{tikzpicture}%
}

\def \vdelta {\bm{\delta}}

\def \Dc {\mathcal{D}}

\newcommand{\ourmethod}{\texttt{Grond}\xspace}
\newcommand{\abi}{ABI\xspace}

\definecolor{upcolor}{rgb}{.1, .75, .1}
\definecolor{downcolor}{rgb}{1., 0., 0.}

\newcommand{\up}[1]{\small \textcolor{gray}{$\uparrow$#1}}
\newcommand{\down}[1]{\small \textcolor{gray}{$\downarrow$#1}}

\definecolor{LightGray}{gray}{0.85}
\definecolor{White}{gray}{1.0}
\definecolor{Celadon}{RGB}{175, 225, 175}

\newcommand{\grayc}{\cellcolor{LightGray}}
\newcommand{\pinkc}{\cellcolor{pink!75}}
\newcommand{\greenc}{\cellcolor{Celadon!75}}

\newcommand{\tabincell}[2]{\begin{tabular}{@{}#1@{}}#2\end{tabular}}

\newcommand\blfootnote[1]{
  \begingroup
  \renewcommand\thefootnote{}\footnote{#1}%
  \addtocounter{footnote}{-1}%
  \endgroup
}

\newcommand{\appcite}[1]{App.~\ref{#1}}

\begin{document}

\title{Towards Backdoor Stealthiness in Model Parameter Space}


\author{Xiaoyun Xu}
\affiliation{%
  \institution{Radboud University}
  \city{Nijmegen}
  \country{The Netherlands}}
\email{xiaoyun.xu@ru.nl}

\author{Zhuoran Liu$^{*}$}
\affiliation{%
  \institution{Radboud University}
  \city{Nijmegen}
  \country{The Netherlands}}
\email{z.liu@cs.ru.nl}

\author{Stefanos Koffas}
\affiliation{%
  \institution{Delft University of Technology}
  \city{Delft}
  \country{The Netherlands}}
\email{s.koffas@tudelft.nl}

\author{Stjepan Picek}
\affiliation{%
  \institution{Radboud University, Nijmegen, The Netherlands}
  \institution{\& University of Zagreb Faculty of Electrical Engineering and Computing, Unska 3, 10000, Zagreb, Croatia}
  \city{}
  \country{}
  }
\email{stjepan.picek@ru.nl}

\renewcommand{\shortauthors}{Xiaoyun Xu et al.}

\begin{abstract}
Backdoor attacks maliciously inject covert functionality into machine learning models, representing a security threat.
The stealthiness of backdoor attacks is a critical research direction, focusing on adversaries' efforts to enhance the resistance of backdoor attacks against defense mechanisms.
Recent research on backdoor stealthiness focuses mainly on indistinguishable triggers in \emph{input space} and inseparable backdoor representations in \emph{feature space}, aiming to circumvent backdoor defenses that examine these respective spaces.
However, existing backdoor attacks are typically designed to resist a specific type of backdoor defense without considering the diverse range of defense mechanisms.
Based on this observation, we pose a natural question: \emph{Are current backdoor attacks truly a real-world threat when facing diverse practical defenses?}
\blfootnote{$^{*}$ Corresponding author.}

To answer this question, we examine 12 common backdoor attacks that focus on input-space or feature-space stealthiness and 17 diverse representative defenses.
Surprisingly, we reveal a critical blind spot that backdoor attacks designed to be stealthy in input and feature spaces can be mitigated by examining backdoored models in \emph{parameter space}. 
To investigate the underlying causes behind this common vulnerability, 
we study the characteristics of backdoor attacks in the parameter space.
Notably, we find that input- and feature-space attacks introduce prominent backdoor-related neurons in parameter space, which are not thoroughly considered by current backdoor attacks.
Taking comprehensive stealthiness into account, we propose a novel supply-chain attack called \ourmethod.
\ourmethod limits the parameter changes by a simple yet effective module, 
Adversarial Backdoor Injection (\abi), which adaptively increases the parameter-space stealthiness during the backdoor injection.
Extensive experiments demonstrate that \ourmethod outperforms all 12 backdoor attacks against state-of-the-art (including adaptive) defenses on CIFAR10, GTSRB, and a subset of ImageNet. 
Additionally, we show that ABI consistently improves the effectiveness of common backdoor attacks.
Our code is publicly available: \url{https://github.com/xiaoyunxxy/parameter_backdoor}.
\end{abstract}




\maketitle

\let\oldclearpage\clearpage 
\let\clearpage\relax     

\begin{table*}[t]
\begin{threeparttable}
    \centering
    \footnotesize
    \caption{A summary of the existing defenses evaluated in this paper. \emph{``Proactively training''} refers to the strategy that the defender could proactively control the training on poisoned training data to produce a clean model without a backdoor in it. Additionally, all the defenses have been tested against the all-to-one attack, so we omitted it from the attack assumptions. A summary of backdoor attacks is provided in Table~\ref{tab:attack_summary} in \appcite{appendix:threat_model}.}
    \label{tab:summary}
    \begin{tabular}{cccccccccccccccccccccccccccccccccc}
        \toprule
        \multicolumn{2}{c}{\multirow{2}{*}{Defense}} & \multicolumn{3}{c}{Defense Task} & \multicolumn{3}{c}{Threat Model} & \multicolumn{2}{c}{Attack Assumption}\\
        \cmidrule(lr){3-5}
        \cmidrule(lr){6-8}
        \cmidrule(lr){9-10}
        ~ & ~ & Input detection & Model detection & Mitigation & Black-box & Needs clean data & Proactively training & All-to-all & Dynamic\\
        \midrule
        \multirow{5}{*}{\tabincell{c}{Model \\ Inspection}} & NC~\cite{wang_2019_nc} & \pie{90} & \pie{360} & \pie{90} & \pie{90} & \pie{360} & \pie{90} & \pie{90} & \pie{90}\\
        ~ & Tabor~\cite{guo_2020_tabor} & \pie{90} & \pie{360} & \pie{90} & \pie{90} & \pie{360} & \pie{90} & \pie{90} & \pie{90}\\
        ~ & FeatureRE~\cite{wang2022featurere} & \pie{90} & \pie{360} & \pie{90} & \pie{90} & \pie{360} & \pie{90} & \pie{90} & \pie{360}\\
        ~ & Unicorn~\cite{wang2023unicorn} & \pie{90} & \pie{360} & \pie{90} & \pie{90} & \pie{360} & \pie{90} & \pie{90} & \pie{360}\\
        ~ & BTI-DBF~\cite{xu2024btidbf} & \pie{90} & \pie{360} & \pie{90} & \pie{90} & \pie{360} & \pie{90} & \pie{360} & \pie{360}\\
        \midrule
        \multirow{3}{*}{\tabincell{c}{Input \\ Inspection}} & Scale-up~\cite{guo2023scaleup} & \pie{360} & \pie{90} & \pie{90} & \pie{360} & \pie{90} & \pie{90} & \pie{360} & \pie{360}\\
        ~ & IBD-PSC~\cite{hou2024ibdpsc} & \pie{360} & \pie{90} & \pie{90} & \pie{90} & \pie{360} & \pie{90} & \pie{360} & \pie{360}\\
        ~ & CT~\cite{qi2023proactivedetection} & \pie{360} & \pie{90} & \pie{90} & \pie{90} & \pie{360} & \pie{360} & \pie{360} & \pie{360}\\
        \midrule
        \multirow{4}{*}{\tabincell{c}{Pruning}} & FP~\cite{liu2018finepruning} & \pie{90} & \pie{90} & \pie{360} & \pie{90} & \pie{360} & \pie{90} & \pie{360} & \pie{360}\\
        ~ & ANP~\cite{Wu_2021_anp} & \pie{90} & \pie{90} & \pie{360} & \pie{90} & \pie{360} & \pie{90} & \pie{360} & \pie{360}\\
        ~ & CLP~\cite{zheng2022clp} & \pie{90} & \pie{90} & \pie{360} & \pie{90} & \pie{90} & \pie{90} & \pie{360} & \pie{360}\\
        ~ & RNP~\cite{li2023rnp} & \pie{90} & \pie{90} & \pie{360} & \pie{90} & \pie{360} & \pie{90} & \pie{360} & \pie{360}\\
        \midrule
        \multirow{5}{*}{\tabincell{c}{Fine-tune}} & vanilla FT & \pie{90} & \pie{90} & \pie{360} & \pie{90} & \pie{360} & \pie{90} & \pie{360} & \pie{360}\\
        ~ & FT-SAM~\cite{Zhu2023ftsam} & \pie{90} & \pie{90} & \pie{360} & \pie{90} & \pie{360} & \pie{90} & \pie{360} & \pie{360}\\
        ~ & I-BAU~\cite{zeng2022ibau} & \pie{90} & \pie{90} & \pie{360} & \pie{90} & \pie{360} & \pie{90} & \pie{360} & \pie{360}\\
        ~ & FST~\cite{min2023fst} & \pie{90} & \pie{90} & \pie{360} & \pie{90} & \pie{360} & \pie{90} & \pie{360} & \pie{360}\\
        ~ & BTI-DBF(U)~\cite{xu2024btidbf} & \pie{90} & \pie{90} & \pie{360} & \pie{90} & \pie{360} & \pie{90} & \pie{360} & \pie{360}\\
        \bottomrule
    \end{tabular}
\begin{tablenotes}
    \item \pie{90} the item is not supported by the defense; \pie{360} the item is supported by the defense.
\end{tablenotes}
\end{threeparttable}
\end{table*}

\section{Introduction}
\label{sec:introduction}

While deep neural networks (DNNs) have achieved excellent performance on various tasks, they are vulnerable to backdoor attacks.
Backdoor attacks insert a secret functionality into a model, which is activated by malicious inputs during inference. Such inputs contain an attacker-chosen property called the trigger.
Backdoored DNNs can be created by training with poisoned data~\cite{Gu_2019_badnet,chen_2017_blend,nguyen2021wanet,turner2019lc}. 
More powerful and stealthy backdoors can also be injected through the control of a training process~\cite{shokri2020bypassing,Bagdasaryan_2021_blind,Cheng2021dfst,nan2022iba,Qi2022deploymentbackdoor,mo2024ssdt,zhao2022defeat}, or by direct weights modification of the victim model~\cite{hong2022handcrafted,cao2024dfba}.

In early backdoor attacks~\cite{Gu_2019_badnet,chen_2017_blend,liu_2018_Trojannn}, triggers could induce noticeable changes that human inspectors or anomaly detectors~\cite{chen2018detecting,wang_2019_nc,liu_2019_abs} could easily spot.
To enhance the ability to remain undetected against such defenses (i.e., achieve \emph{input-space stealthiness}), 
smaller or more semantic-aware triggers are designed~\cite{nguyen2021wanet,Doan2021lira,Wang_2022_bpp}.
Input-space stealthy backdoor attacks usually need to change labels of poisoned samples to the target class (i.e., dirty-label), which makes detection easier~\cite{chen2018detecting}.
To this end, another line of backdoor attacks poisons the training data without changing the labels~\cite{turner2019lc,zeng2023narcisus} (i.e., clean-label), improving backdoor stealthiness.

Despite the stealthiness concerning input images and labels, it has been widely observed that existing backdoor attacks introduce separable representations in the feature space, which can be exploited to develop backdoor defenses~\cite{wang2022featurere,qi2022revisitingadaptive,min2023fst,Zhu2023ftsam,xu2024btidbf}.
For example, featureRE~\cite{wang2022featurere} utilizes feature separability and designs a feature space constraint to reverse-engineer the backdoor trigger.
In response to feature-space defenses, state-of-the-art (SOTA) backdoor attacks focus on eliminating the separability in the feature space~\cite{mo2024ssdt,qi2022revisitingadaptive,abad2024gradientshaping,shokri2020bypassing} to increase the \emph{feature-space stealthiness}, i.e., the undetectability against feature-space defenses.
Considering a different threat model, supply-chain backdoor attacks assume control over the training or directly modify the model's weights~\cite{Cheng2021dfst,hong2022handcrafted,cao2024dfba}, and the backdoored model is provided as a service or as the final product.
For example, supply-chain attacks could introduce a penalty to the training loss that decreases the distance between the backdoor and benign features to increase feature-space stealthiness~\cite{shokri2020bypassing,zhao2022defeat,doan2021wb,nan2022iba}.

An important observation is that most backdoor attacks are designed to be stealthy to resist a specific type of defense.
For example, WaNet~\cite{nguyen2021wanet} and Bpp~\cite{Wang_2022_bpp} design imperceptible triggers to bypass input-space defenses (such as NC~\cite{wang_2019_nc}), but introduce significant separability in the feature space~\cite{wang2022featurere}.
Adap-patch~\cite{qi2022revisitingadaptive} avoids feature separability but uses patch-based triggers, which a human inspector can detect.
More critically, current backdoor attacks are barely evaluated against \emph{parameter-space defenses}~\cite{Zhu2023ftsam,lin2024unveiling,xu2024ban, Wu_2021_anp,zheng2022clp,li2023rnp}. 
This oversight is significant because backdoor behaviors are ultimately embedded in and reflected by the parameters of the backdoored model, which is the final product of any backdoor attack. 
As such, there is a lack of systematic evaluation of backdoor attacks against the latest \emph{parameter-space defenses}.

To this end, in this paper, we first systematically analyze 12 attacks against 17 backdoor defenses. 
All evaluated defenses and their characteristics, including detection and mitigation, are summarized in Table~\ref{tab:summary}. 
Surprisingly, our experiments demonstrate that parameter-space defenses can easily mitigate SOTA stealthy backdoor attacks (including supply-chain attacks),
indicating that existing stealthy backdoor attacks fail to provide \emph{parameter-space stealthiness} and, as a result, still need substantial improvement to be stealthy in the model's parameter space.
More importantly, our analysis reveals that even though some backdoor attacks can resist several defenses, bypassing all defense types is far from trivial.

\begin{figure*}[t]
\centering
\includegraphics[width = 0.9\linewidth]{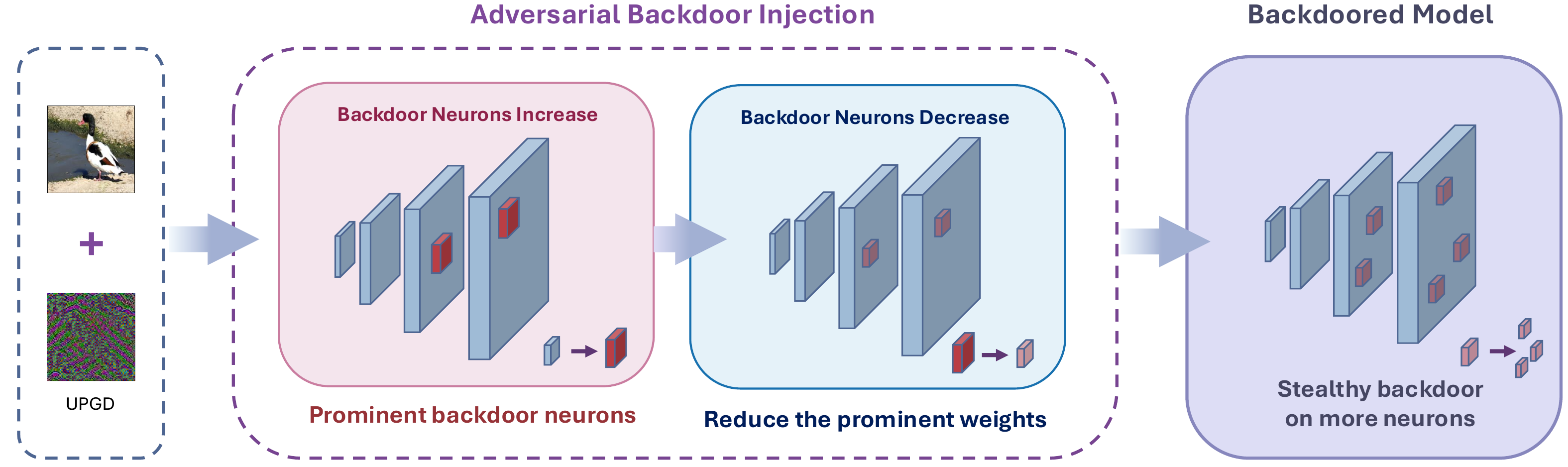}
\caption{Diagram illustrating the working mechanism of \ourmethod. 
On the left, universal PGD (UPGD) perturbation is generated as backdoor patterns to be injected. 
In the middle, \abi is applied where perturbed samples are iteratively used to train the model, and the model parameters are pruned to limit the magnitude of prominent backdoored weights. 
On the right, the output backdoored model that considers comprehensive stealthiness is deployed, where 1) the triggers are invisible, 2) the features of trigger samples are inseparable, and 3) the backdoored model weights are hardly distinguishable from benign model weights. 
Perturbations generated by UPGD are scaled up $10\times$ for visualization.}
\label{fig:method}
\end{figure*}

To explore whether it is possible to make backdoor attacks stealthy simultaneously against diverse defenses, we propose a novel attack called \ourmethod 
that considers \emph{comprehensive stealthiness}, meaning that a backdoor attack is stealthy in the input, the feature, and the parameter space of the model.
\ourmethod achieves the input space stealthiness by using adversarial perturbations as the trigger.
To achieve parameter-space stealthiness, we propose a novel \emph{Adversarial Backdoor Injection} module that adaptively injects the backdoor during the backdoor training to achieve parameter space stealthiness.
We also show that the feature-space stealthiness is a by-product of input- and parameter-space stealthiness with empirical results in Figures~\ref{fig:feature_tsne} and~\ref{fig:featureloss}. 
Specifically, guided by our TAC analysis, we leverage the Lipschitz continuity of neuron activations to find backdoor-related suspicious and sensitive neurons.
Then, we conduct pruning on these neurons to eliminate the backdoor effect.
As a result, the backdoor is associated with neurons throughout the DNN rather than just focusing on a few prominent neurons after Adversarial Backdoor Injection, as illustrated in Figure~\ref{fig:method}.

We make the following contributions:
\begin{compactitem}
    \item We revisit SOTA backdoor attacks regarding their stealthiness, showing that most attacks are designed to increase input-space indistinguishability or/and feature-space inseparability without considering parameter-space stealthiness.
    Based on this finding, we examine common backdoor attacks and reveal a critical \emph{blind spot} regarding real-world scenarios: SOTA stealthy backdoor attacks are highly vulnerable to parameter-space defenses.
    \item To investigate the underlying reasons behind this common vulnerability of backdoor attacks, we take a closer look at the backdoor characteristics in the parameter space, showing that input- and feature-space attacks introduce prominent backdoor-related neurons, which cannot be avoided by current backdoor attacks.
    \item To accomplish comprehensive stealthiness, we propose a novel backdoor attack, \ourmethod, that considers input, feature, and parameter-space defenses.
    Extensive experiments demonstrate that \ourmethod outperforms SOTA attacks against four pruning- and five fine-tuning-based defenses on CIFAR10, GTSRB, and ImageNet200. 
    Moreover, we demonstrate that \ourmethod is resistant against five model detection defenses, two input detection defenses, and a proactive defense.
    \item We verify the effectiveness of the Adversarial Backdoor Injection module by binding it with other attacks. 
    Experimental results demonstrate that Adversarial Backdoor Injection could substantially improve the parameter-space robustness of most common backdoor attacks.     
\end{compactitem}

\section{Background \& Related Work}

\subsection{Preliminaries on Backdoor Training}
\label{sec:relatedwork-pre}

This paper considers a $C$-class classification problem with an $L$-layer CNN $f=f_L\circ \cdots f_1$.
Suppose that $\Dc = \{ (\vx_i, y_i) \}_{i=1}^{N}$ is the original training data, containing $N$ samples of $\vx_i \in \R^{d_c \times d_h \times d_w}$ and its label $y \in \{1, 2,\ldots, C\}$.
$d_c$, $d_h$, and $d_w$ are the number of input channels, the height, and the width of the image, respectively.
The attacker chooses a target class $t$ and creates a partially poisoned dataset $\Dc_p$ by poisoning generators $G_x$ and $G_y$, i.e., $\Dc_p = \Dc_c \cup \Dc_b$.
$\Dc_c$ is the benign data from original dataset, $\Dc_b = \{ (\vx',y') | \vx' = G_x(\vx), y' = G_y(y), (\vx, y) \in \Dc - \Dc_c \}$.
In the clean-label setting, $G_y(y) = y$. For the dirty-label attacks, $G_y(y) = t$.
In the training stage, the backdoor is inserted into $f$ by minimizing the loss on $\Dc_p$:
\begin{equation}
    \mathop{\min}\limits_{\vtheta} \mathcal{L}_{\mathcal{D}_p} (\vtheta) = \mathop{\mathbb{E}}\limits_{(\vx, y) \in \mathcal{D}_p} \ell (f(\vx; \vtheta), y).
\end{equation}
In the inference stage, the trained $f$ performs well on benign data $\hat{\vx}$, but predicts $G_x(\hat{\vx})$ as $G_y( \hat{y} )$.

\subsection{Backdoor Attacks}
\label{sec:relatedwork-BA}

\begin{figure*}[htb]
\centering
\includegraphics[width = 1\linewidth]{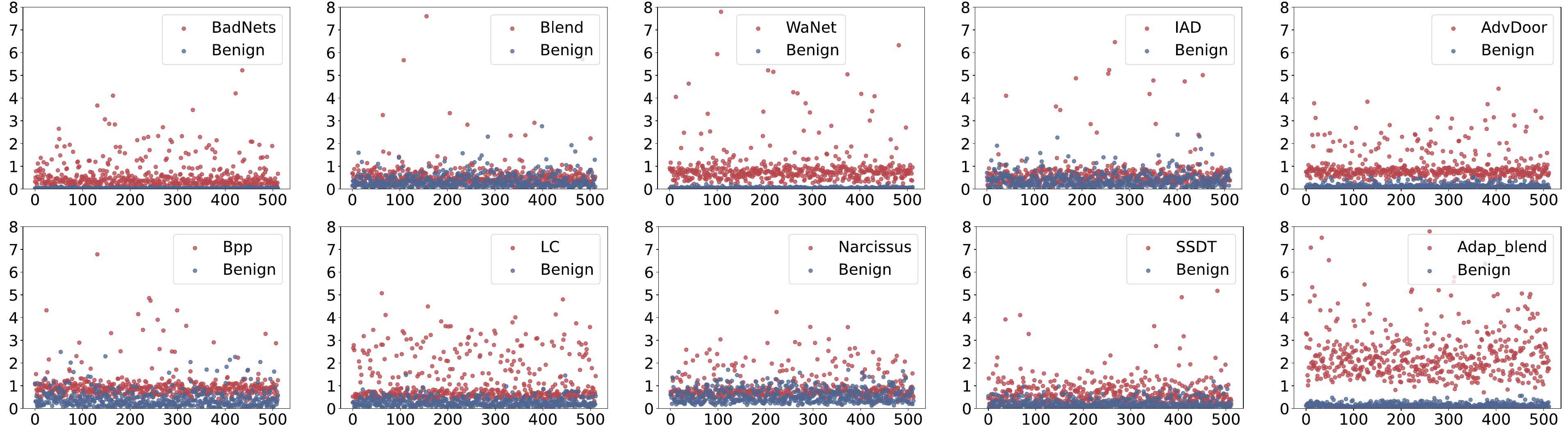}
\caption{TAC~\cite{zheng2022clp} analysis of different backdoor attacks on 512 neurons. The $y$ axis contains the TAC values, and the $x$ axis depicts the index of neurons.
Higher TAC values suggest a stronger relation between corresponding neurons and the backdoor trigger.
In the Appendix~\appcite{appendix:tac_others}, we also provide sorted TAC value plots, clearly showing the prominent TAC values.
}
\label{fig:tac_all_baseline}
\end{figure*}

Backdoor attacks compromise the integrity of the victim model so that the model performs naturally on benign inputs but is misled to the target class by inputs containing the backdoor trigger.
The trigger can be a visible pattern inserted into the model's input in the \textbf{input space} or a property that affects the feature representation of the model's input in the \textbf{feature space}. 
Eventually, however, the backdoored model's parameters in the \textbf{parameter space} will be altered regardless of the exact backdoor attack (see Figure~\ref{fig:tac_all_baseline}).
To insert a backdoor, the attacker is assumed to only control a small portion of the training data under the \emph{poison training} scenario~\cite{Gu_2019_badnet,chen_2017_blend,zhang2021advdoor}.
In the \emph{supply-chain} setting (backdoor models provided to users),
the attacker also controls the training process~\cite{shokri2020bypassing,nguyen2020inputaware,Bagdasaryan_2021_blind,nguyen2021wanet,Wang_2022_bpp}.
Moreover, the backdoor can also be created by directly modifying the model's weights~\cite{liu2017faultinjection,hong2022handcrafted,Qi2022deploymentbackdoor,cao2024dfba}.

\noindent
\textbf{Input-space attacks.} Traditional attacks typically use simple patterns as their triggers.
For example, BadNets~\cite{Gu_2019_badnet} uses a fixed patch, and Blend~\cite{chen_2017_blend} mixes a Hello Kitty pattern into the images as the trigger.
These non-stealthy triggers introduce abnormal data into training data and can be easily detected by human inspectors or defenses~\cite{chen2018detecting,wang_2019_nc}.
To improve the stealthiness, various triggers are proposed to achieve \emph{invisibility} in the input space.
IAD~\cite{nguyen2020inputaware} designed a dynamic solution in which the triggers vary among different inputs.
WaNet~\cite{nguyen2021wanet} proposed the warping-based trigger, which is invisible to human inspection.
Although these methods successfully build invisible triggers and bypass traditional defenses~\cite{wang_2019_nc}, they still introduce separable features and can be detected by feature-space defenses~\cite{wang2022featurere, xu2024btidbf}.
These input-invisible attacks can be even more noticeable than input-visible attacks (BadNet, Blend) in the feature space~\cite{xu2024ban}. 
We conjecture this is because they have fewer modifications on input pixels than input-visible attacks.
Therefore, input-invisible attacks require more influential features to achieve a successful attack.

\noindent
\textbf{Feature-space attacks.} 
Knowing the vulnerability of input-space attacks against feature-space defenses, backdoor attacks are improved for feature-space stealthiness.
A common threat model of this attack type is to assume additional control over the training process.
For example,~\cite{shokri2020bypassing,doan2021wb,zhao2022defeat,nan2022iba} directly designed loss functions to minimize the difference between the backdoor and benign features.
Aside from design loss penalties, TACT~\cite{di2021tact} and SSDT~\cite{mo2024ssdt} point out that source-specific (poison only the specified source classes) attack helps to obscure the difference in features between benign and backdoor samples.
In addition,~\cite{qi2022revisitingadaptive} proposed Adap-blend and Adap-patch, which obscures benign and backdoor features by 1) including poisoned samples with the correct label, 2) asymmetric triggers (using a stronger trigger at inference time), and 3) trigger diversification (using diverse variants of the trigger during training).
Unfortunately, existing attacks lack systematic evaluation against the latest defenses.
For example, Adap-blend can be thoroughly mitigated by recent works~\cite{Zhu2023ftsam,xu2024btidbf,xu2024ban}.
In summary, feature-space attacks usually introduce visible triggers and cannot defeat the latest defenses.

\noindent
\textbf{Supply-chain attacks.}
Supply-chain attacks are getting more attention due to their potential in real-world applications where backdoored models are provided as the final product to users. 
In supply-chain attacks, adversaries could control both training data and the training process.
Note that feature-space attacks~\cite{shokri2020bypassing,liu2020compositeattack,doan2021wb,zhao2022defeat,ren2021Simtrojan,Cheng2021dfst,nan2022iba,xia2023mmdregularization,mo2024ssdt} with the assumption of control over the training process are a subset of supply-chain attacks, as their output is the backdoor model.
In addition to training control, another kind of supply-chain attack directly adjusts the model's weights in parameter space to introduce a backdoor, i.e., \emph{parameter-space attack}.
T-BFA~\cite{rakin2022tbfa}, TBT~\cite{Rakin_2020_TBT}, and ProFlip~\cite{chen2021proflip} explore modifying a sequence of susceptible bits of DNN parameters stored in the main memory (e.g., DRAM) to inject the backdoor.
SRA~\cite{Qi2022deploymentbackdoor} and handcrafted Backdoor~\cite{hong2022handcrafted} directly modify a subset of models' parameters to increase the logits of the target class.
However, these attacks require a local benign dataset to guide the search for the subset of parameters to be modified.
Data-free backdoor~\cite{lv2023datafree} releases the requirement of benign data by collecting substitute data irrelevant to the main task and fine-tuning using the substitute data.
DFBA~\cite{cao2024dfba} further proposes a retraining-free and data-free backdoor attack by injecting a backdoor path (a single neuron from each layer except the output layer) into the victim model.
In summary, supply-chain attacks focus on increasing the backdoor's effectiveness without comprehensively considering parameter-space defenses.

\subsection{Backdoor Defenses}
\label{sec:relatedwork-BD}

Backdoor defenses can be classified into detection and mitigation.
Detection refers to determining whether a model is backdoored (\emph{model detection})~\cite{wang_2019_nc,liu_2019_abs,zhao2022defeat,wang2023unicorn,xu2024btidbf} or a given input is applied with a trigger (\emph{input detection})~\cite{gao2019strip,guo2023scaleup,mo2024ssdt}.
Mitigation refers to erasing the backdoor effect from the victim model by pruning the backdoor-related neurons (\emph{pruning-based} defenses)~\cite{liu2018finepruning,Wu_2021_anp,zheng2022clp,li2023rnp} or unlearning the backdoor trigger (\emph{fine-tuning-based} defenses)~\cite{Zhu2023ftsam,zeng2022ibau,min2023fst,xu2024btidbf}.
In addition, recent works~\cite{li2021abl,qi2023proactivedetection,wei2024pdb} also consider the home-field advantage\footnote{The defender has full control of the system and could access the training process.} to design more powerful \emph{proactive defenses}.

\noindent
\textbf{Backdoor detection.}
Backdoor trigger reverse engineering (also known as trigger inversion) is considered one of the most practical defenses for backdoor detection as it can be applied to both poisoning training and supply-chain scenarios~\cite{wang2022featurere,wang2023unicorn,xu2024btidbf,xu2024ban}, i.e., it is a post-training method.
Specifically, trigger inversion works by searching for a potential backdoor trigger for a specific model.
The model is determined as backdoored if a trigger is found, and the trigger can be used to unlearn the backdoor.
The searching is implemented as an optimization process corresponding to the model and a local benign dataset.
For example, NC~\cite{wang_2019_nc} firstly proposes trigger inversion for detection by optimizing the mask and pattern in the input space that can mislead the victim to the target class.
This optimization is repeated for all classes.
The model is considered backdoored if an outlier significantly smaller than the triggers for all other classes exists.
Although methods similar to NC perform well against fixed patch trigger attacks, such as BadNets~\cite{Gu_2019_badnet} and Blend~\cite{chen_2017_blend}, they may not be effective against input-stealthy attacks like WaNet~\cite{nguyen2021wanet}.
To address this problem, FeatureRE~\cite{wang2022featurere} moves trigger inversion from input space to feature space.
Unicorn~\cite{wang2023unicorn} further proposes a transformation function for attacks in other spaces, such as numerical space~\cite{Wang_2022_bpp}.
Recent works~\cite{xu2024btidbf,xu2024ban} focus on exploring new optimization objectives that address the inefficiency problem of previous trigger inversion methods due to optimization over all classes.
BTI-DBF~\cite{xu2024btidbf} trains a trigger generator by maximizing the backdoor feature difference between benign samples and their generated version (by the trigger generator) and minimizing the benign feature difference.
BAN~\cite{xu2024ban} optimizes the noise on neuron weights rather than input pixels to activate the potential backdoor, which further improves both effectiveness and efficiency.

\noindent
\textbf{Backdoor mitigation.}
Backdoor mitigation consists of fine-tuning and pruning, which are effective and do not assume knowledge of backdoor triggers.
Pruning methods aim to find and remove backdoor-related neurons.
FP~\cite{liu2018finepruning} eliminates dormant neurons on benign inputs and then fine-tunes the pruned network.
ANP~\cite{Wu_2021_anp} searches for backdoor-related neurons by adding adversarial noise to neuron weights to activate the backdoor.
RNP~\cite{li2023rnp} uses an unlearning and recovering process on benign data to expose backdoor neurons, as the recovering will force the backdoor neurons to be silent for the main benign task.
Unlike these pruning methods guided by benign data, CLP~\cite{zheng2022clp} directly analyzes the Channel Lipschitzness Constant of the network and prunes the high Lipschitz constant channels in a data-free manner.

Traditional fine-tuning as a defense usually needs trigger inversion methods to recover the trigger and then unlearn the trigger.
For example, BTI-DBF(U)~\cite{xu2024btidbf} fine-tunes backdoor models using triggers recovered by their inversion algorithm.
However, there is no guarantee that the recovered trigger is the true trigger for the backdoor.
Recent works also consider fine-tuning without the trigger information but with prior human knowledge.
For example, FT-SAM~\cite{Zhu2023ftsam} observes a positive correlation between the weight norm of neurons and backdoor-related neurons.
Then, they propose a fine-tuning method to revise the large outliers of weight norms using Sharpness-Aware Minimization (SAM).
I-BAU~\cite{zeng2022ibau} forms a min-max fine-tuning similar to adversarial training, where the inner maximizing searches for perturbations that mislead the model, and the outer minimizing is to keep the model's capability on benign data.
FST~\cite{min2023fst} assumes the backdoor and benign features should be disentangled and actively shifting features while fine-tuning by encouraging the discrepancy between the original backdoor model and the fine-tuned model. 

\noindent
\textbf{Proactive defense.}
Several methods have been proposed to exploit the home-field advantage, i.e., a stronger defender, for better defensive performance.
ABL~\cite{li2021abl} proposes two techniques to avoid learning the backdoor task while training on the poisoned data: 1) trapping the loss value of each example around a certain threshold because backdoor tasks are learned much faster than the main task, and their loss decreases much faster. The samples with lower loss are recorded as poisoned samples; 2) unlearning the backdoor with the recorded poisoned samples.
CT~\cite{qi2023proactivedetection} detects poisoned samples in the training set by introducing confusing batches of benign data with randomly modified labels.
The confusing batches with random labeling corrupt the benign correlations between normal semantic features and semantic labels, so the inference model trained with confusing batches and the poisoned dataset will find it hard to distinguish benign samples.
However, the correlation between the backdoor trigger and the target label remains intact, as the confusing batches contain no trigger.
Therefore, samples with correctly predicted labels by the inference model are considered poisoned.
PDB~\cite{wei2024pdb} proactively injects a defensive backdoor into the model during training, overriding the potential backdoor injected by the poisoned training data.
In summary, proactive defenses assume a stronger defender for better defensive performance.

\section{Comprehensive Backdoor Stealthiness}
\label{sec:method}

\subsection{Threat Model}
\label{sec:theatmodel}

\noindent
\textbf{Attacker's goal.} 
The attacker provides pre-trained models to users. The aim is to inject backdoors into the pre-trained model so that the model performs well on clean inputs but predicts the attacker-chosen target label when receiving inputs with a backdoor trigger, i.e., an all-to-one attack.

\noindent
\textbf{Attacker's knowledge.}
The attacker has white-box access to the training processes, the training data, and the model weights, i.e., the supply-chain threat model. 
During inference, the backdoor trigger is imperceptible to human inspectors.

\noindent
\textbf{Attacker's capabilities.}
The attacker can train a well-performed surrogate model to generate UPGD, which is used to perturb the victim model's input. 
Additionally, the attacker can alter the model's weights during training. 
Table~\ref{tab:attack_summary} in \appcite{appendix:threat_model} shows that the threat model of \ourmethod is aligned with baseline attacks.

\subsection{Lack of Parameter-Space Stealthiness}
\label{sec:lack}

As introduced in the related work, early backdoor attacks that introduce noticeable changes in either input~\cite{Gu_2019_badnet,chen_2017_blend} or feature space~\cite{nguyen2021wanet,nguyen2020inputaware} have been empirically shown powerful, even with very low poisoning rates~\cite{Gu_2019_badnet,zeng2023narcisus}.
Focusing on the backdoor-introduced noticeable changes, backdoor defenses are improved to distinguish backdoor patterns in either input or feature space~\cite{wang2022featurere,lin2024unveiling}. 
Meanwhile, backdoor attacks are optimized to increase stealthiness in input~\cite{nguyen2021wanet} or feature space~\cite{qi2022revisitingadaptive}. 
However, regardless of the implementation of input- or feature-space attack logic, backdoor behaviors are eventually embedded in the backdoored model's parameters. 
For this reason, it is important to investigate whether backdoor attacks introduce visible changes in the parameter space of the attacked models that can be used by the parameter-space defenses.

Considering this observation, we ran an initial experiment to understand the behavior of neurons in backdoored models. We use the TAC values~\cite{zheng2022clp} to quantify the relevance of a neuron to the backdoor behavior.
The TAC values show the change in the output of each neuron (a feature channel of convolutional layers) before and after the trigger is attached to the input. 
Thus, TAC quantifies a neuron's sensitivity to the backdoor trigger. 
A high TAC value indicates that the neuron is strongly related to the backdoor behavior, whereas a low TAC value shows it is not. 
TAC takes the exact trigger information into account when measuring the backdoor effect, which makes TAC a straightforward and effective method that captures the relevant backdoor neurons' behaviour.
Specifically, TAC is defined as:
\begin{equation}
    \text{TAC}_l^{(k)}(\Dc_c) = \frac{1}{|\Dc_c|} \sum_{\vx \in \Dc_c} || f_l^{(k)}(\vx) - f_l^{(k)}( G_x (\vx) ) ||_2,
\end{equation}
where $f_l^{(k)}$ is the $k_{th}$ channel of the $l_{th}$ layer.
$G_x (\vx)$ is the poisoned sample.
$\Dc_c$ consists of a few benign samples.
Note that TAC can only be used to analyze backdoor behaviors and cannot be deployed as a practical defense, as it requires access to backdoor triggers, which is unrealistic in practice.

TAC analysis of different backdoor attacks is shown in Figure~\ref{fig:tac_all_baseline}, where each dot represents the TAC value for one of the 512 individual neurons.
We can observe that the TAC values of neurons of backdoored models are substantially higher than those of benign models. 
In particular, neurons with higher TAC values contribute more to the backdoor behavior.
The working mechanism of pruning- and fine-tuning-based backdoor defenses can be understood as targeting and eliminating neurons with TAC values that are substantially higher than those of others.
Our observations from the TAC analysis suggest that backdoor attacks are designed to be stealthy in input space, and feature space can, in fact, be identified in parameter space, making them susceptible to parameter-space defenses. 
Our experimental analysis further substantiates this assumption (see Section~\ref{sec:experiments}). 
Thus, we conclude that current backdoor attacks may not be robust against parameter-space defenses.

\subsection{\ourmethod for Comprehensive Stealthiness}
\label{sec:grond}

To address the vulnerabilities identified in the parameter space, we propose a stealthy backdoor attack, \ourmethod, that considers comprehensive stealthiness, i.e., stealthiness in input, feature, and parameter space.
\ourmethod includes two key parts: UPGD trigger generation and Adversarial Backdoor Injection (ABI). 

\noindent
\textbf{Backdoor trigger generation for input-space stealthiness.} 
We use imperceptible adversarial perturbations to generate \emph{imperceptible} backdoor triggers inspired by adversarial example studies~\cite{Moosavi2017uap,zhang2021advdoor}. 
We modify the original PGD algorithm to generate a universal PGD (UPGD) perturbation as the backdoor trigger.
UPGD contains non-robust but generalizable semantic information~\cite{tsipras2018robustnessoddwithacc}, which correlates with the benign functions of the victim model and shortens the distance between poisoned data and the target classification region~\cite{zhang2021advdoor}.
Consequently, backdoor patterns tend to make fewer prominent changes to the victim network.

Similar to~\cite{zeng2023narcisus,zhang2021advdoor}, UPGD is generated on a well-trained surrogate model trained on the clean training set. 
The architecture and parameters of the surrogate model do not necessarily need to be the same as the victim model (see Table~\ref{tab:architectures} in \appcite{sec:diff_arch_surrogate}).
UPGD is optimized following the PGD~\cite{madry2018pgd} algorithm to decrease the surrogate model's cross-entropy loss that takes as inputs the adversarial examples (the poisoned samples in our case) and the target class label. 
This procedure is described formally in Algorithm~\ref{alg:upgd}.
The $\vdelta$ is the generated UPGD that will be used as a backdoor trigger; thus, $G_x(\vx) = \vx + \vdelta$. 
$S$ is the ball function with the radius $\epsilon$, and the small $\epsilon$ guarantees the imperceptibility of the backdoor trigger as it controls the perturbation's magnitude.

The backdoor is injected during training by poisoning some training data from the target class, i.e., applying the UPGD trigger to the training data.
In the inference stage, our backdoor is activated by the same trigger.
The motivation for our small-size trigger ($\epsilon=8$) is imperceptibility.

\renewcommand{\algorithmicrequire}{\textbf{Input:}}
\renewcommand{\algorithmicensure}{\textbf{Output:}}
\begin{algorithm}[t]
	\caption{UPGD Generation Algorithm} 
	\label{alg:upgd} 
	\begin{algorithmic}[1]
		\REQUIRE{Surrogate model $f_{\theta_{sur}}$, training data $\Dc$, perturbation budget $\epsilon$, the number of iteration $\mathcal{I}$, the target class $t$.}
		\ENSURE{UPGD $\vdelta$}
            \STATE $S = B(\vdelta; \epsilon) = \{ \vdelta \in \mathbb{R}^{d_c \times d_h \times d_w}  : ||\vdelta||_{\infty} \leq \epsilon \}$
            \STATE $\vdelta \leftarrow $ \texttt{random\_initialization} $\land \vdelta \in S$
		\FOR{$i \in (0, \mathcal{I}-1)$}
		\STATE $\vx \leftarrow$ \texttt{sample\_batch($\Dc$)}
            \STATE $\mathcal{L}_{\Dc} (\vtheta) = \mathop{\mathbb{E}}\limits_{(\vx, y) \in \Dc} \ell (f_{\theta_{sur}}(\vx + \bm{\delta}; \vtheta), t), $
            \STATE $\vdelta \leftarrow \mathop{\min}\limits_{\vdelta \in S} \mathcal{L}_{\Dc} (\vtheta)$
		\ENDFOR 
	\end{algorithmic} 
\end{algorithm}

\noindent
\textbf{Adversarial Backdoor Injection for parameter-space stealthiness.}  
Backdoor neurons (i.e., trigger-related neurons) regularly show higher activation values for inputs that contain the trigger, which results in powerful performance~\cite{liu_2019_abs,wang2022featurere,lin2024unveiling}.
To this end, backdoor training needs to substantially increase the magnitude of parameters of backdoor neurons~\cite{Wu_2021_anp,li2023rnp,zheng2022clp}, which harms the parameter-space stealthiness of backdoor attacks. 

One way to find the sensitive neurons with higher activation values is to analyze the Lipschitz continuity of the network.
Leveraging this fact, we introduce a novel backdoor training mechanism, \emph{Adversarial Backdoor Injection}, to increase the parameter-space backdoor stealthiness. 
Specifically, each neuron's Upper bound of Channel Lipschitz Condition (UCLC~\cite{zheng2022clp}) is calculated, based on which the weights of these suspicious neurons are set to the mean of all neurons' weights in the corresponding layer after every training epoch.
In our implementation, we use the weights before every batch normalization as the neuron weights corresponding to the channel setting in UCLC.
We prune neurons by substituting their weights with the mean ones because pruning to zeros makes the training unable to converge in our experiments.
Formally, the $k_{th}$ parameter of the $l_{th}$ layer, $\vtheta_l^{(k)}$, is updated as follows:
\begin{equation}
\label{eq:advesarial_injection}
    \begin{aligned}
            \vtheta_l^{(k)} := \left\{ 
            \begin{aligned}
                & \text{mean}(\vtheta_l), &\sigma(\vtheta_l^{(k)}) > \text{mean}(\sigma(\vtheta_l)) + u \times \text{std}(\sigma(\vtheta_l))\\
                & \vtheta_l^{(k)}, &\text{otherwise,}
            \end{aligned}
            \right.
    \end{aligned}
\end{equation}
where $u$ is a fixed threshold and $\sigma$ is the UCLC value of the given weights.
The measure for quantifying backdoor relevance can be changed from UCLC to others, such as the distance of neuron outputs when receiving benign and backdoor inputs, where a larger distance means the neuron is more relevant to backdoor behaviors and can be pruned.
We use the modified UCLC for training efficiency, as UCLC is data-free, which does not require calculation based on the outputs of neurons.

In adversarial training~\cite{madry2018pgd}, adversarial examples are introduced during training to increase the model's robustness during inference. Similarly, during the Adversarial Backdoor Injection, we use backdoor defenses to increase the resistance of backdoor attacks to parameter-space defenses.
At the end of each training epoch, Adversarial Backdoor Injection prunes the trained model to decrease the weights of backdoor neurons.
Iteratively, backdoored neurons spread across the whole model instead of forming a few prominent backdoor neurons, as illustrated in Figure~\ref{fig:method}.

\noindent
\textbf{Feature-space stealthiness.}
We hypothesize that feature-space stealthiness is a by-product of parameter-space and input-space stealthiness since the variation of feature maps is strongly correlated with model parameters and inputs. 
Figures~\ref{fig:feature_tsne} and~\ref{fig:featureloss} show that \ourmethod can substantially increase the feature-space stealthiness.

\begin{table*}[tb]
\centering
\caption{Pruning-based mitigations against backdoored ResNet18 on CIFAR10. BA refers to benign accuracy on clean data, ASR to attack success rate, and PR to the poisoning rate of the training set. The average drop of BA and ASR is also shown with downward arrows compared to the performance without any defense. Red marks indicate the attack failed to resist the defense with an ASR lower than 60\%, and green means that the ASR is higher than 60\%.}
\label{tab:pruning}
\renewcommand{\arraystretch}{0.95}
\begin{tabular}{ccccccccccccc}
\toprule
\multirow{2}{*}{Attack} & \multicolumn{2}{c}{No Defense} & \multicolumn{2}{c}{FP~\cite{liu2018finepruning}} & \multicolumn{2}{c}{ANP~\cite{Wu_2021_anp}} & \multicolumn{2}{c}{CLP~\cite{zheng2022clp}} & \multicolumn{2}{c}{RNP~\cite{li2023rnp}} & \multicolumn{2}{c}{Average}\\
\cmidrule(lr){2-3}
\cmidrule(lr){4-5}
\cmidrule(lr){6-7}
\cmidrule(lr){8-9}
\cmidrule(lr){10-11}
\cmidrule(lr){12-13}
~ & BA & ASR & BA & ASR & BA & ASR & BA & ASR & BA & ASR & BA & ASR\\
\midrule
BadNets~\cite{Gu_2019_badnet} & 93.13 & \grayc 100 & 92.42 & \greenc 71.71 & 91.60 & \pinkc 1.06 & 88.99 & \pinkc 49.02 & 84.04 & \pinkc 13.82 &  89.26 \down{3.87} & \pinkc33.90 \down{66.10} \\
Blend~\cite{chen_2017_blend}  & 94.42 & \grayc 100 & 93.08 & \greenc 99.99 & 93.57 & \pinkc 0.33 & 90.3 & \pinkc 0.54 & 94.63 & \pinkc 57.98 & 92.89 \down{1.53} & \pinkc 39.71 \down{60.29}\\
WaNet~\cite{nguyen2021wanet} & 93.60 & \grayc 99.37 & 92.96 & \pinkc 4.60 & 91.08 & \pinkc 0.49 & 91.53 & \pinkc 2.12 & 92.86 & \pinkc 3.17 & 92.11 \down{1.49} & \pinkc 2.59 \down{96.78}\\
IAD~\cite{nguyen2020inputaware} & 92.88 & \grayc 97.10 & 91.96 & \pinkc 1.22 & 92.84 & \pinkc 0.71 & 92.24 & \pinkc 0.74 & 92.72 & \pinkc 0.42  & 92.44 \down{0.44} & \pinkc 0.77 \down{96.33}\\
AdvDoor~\cite{zhang2021advdoor} & 93.97 & \grayc 100 & 93.37 & \greenc 98.69 & 91.46 & \pinkc 28.83 & 89.22 & \pinkc 6.13 & 90.17 & \pinkc 44.60 & 91.05 \down{2.92} & \pinkc 44.56 \down{55.44}\\
Bpp~\cite{Wang_2022_bpp} & 94.19 & \grayc 99.93 & 93.38 & \pinkc 18.89 & 92.96 & \pinkc 2.97 & 93.37 & \pinkc 1.89 & 92.2 & \pinkc 5.79 & 92.98 \down{1.21} & \pinkc 7.39 \down{92.54}\\
LC~\cite{turner2019lc} & 94.31 & \grayc 100 & 92.22 & \greenc 93.57 & 91.02 & \pinkc 24.43 & 90.96 & \pinkc 0.38 & 82.70 & \pinkc 33.60 & 89.23 \down{5.08} & \pinkc 37.99 \down{62.01}\\
Narcissus~\cite{zeng2023narcisus} & 93.58 & \grayc 99.64 & 93.49 & \greenc 96.54 & 89.76 & \pinkc 49.18 & 93.19 & \greenc 97.82 & 91.10 & \greenc 94.59 & 91.88 \down{1.70} & \greenc 84.53 \down{15.11}\\
Adap-blend~\cite{qi2022revisitingadaptive} & 92.74 & \grayc 99.67 & 92.06 & \greenc 95.50 & 86.48 & \greenc 67.73 & 92.49 & \greenc 99.62 & 78.63 & \pinkc 1.56 & 87.42 \down{5.32} & \greenc 66.10 \down{33.57}\\
SSDT~\cite{mo2024ssdt} & 93.70 & \grayc 90.30 & 93.41 & \pinkc 0.80 & 93.88 & \pinkc 0.60 & 93.66 & \pinkc 1.20 & 93.99 & \pinkc 3.30 & 93.74 \up{0.04} & \pinkc 1.47 \down{88.83}\\
DFST~\cite{Cheng2021dfst} & 95.23 & \grayc 100 & 94.79 & \greenc 93.84 & 94.64 & \pinkc 3.72 & 92.43 & \pinkc 3.53 & 93.29 & \pinkc 12.68 & 93.79 \down{1.44} & \pinkc 28.44 \down{71.56}\\
DFBA~\cite{cao2024dfba} & 88.99 & \grayc 100 & 86.85 & \pinkc 0.03 & 88.96 & \pinkc 9.55 & 88.96 & \pinkc 9.57 & 88.96 & \pinkc 0.90 & 88.43 \down{0.56} & \pinkc 5.01 \down{94.99}\\
\midrule
\ourmethod (PR=5\%) & 93.43 & \grayc 98.04 & 93.09 & \greenc 99.73 & 91.43 & \greenc 94.01 & 93.29 & \greenc 87.89 & 91.83 & \greenc 85.22 & 92.41 \down{1.02} & \greenc 91.71 \down{6.33}\\
\ourmethod (PR=1\%) & 94.26 & \grayc 93.51 & 93.31 & \greenc 96.32 & 92.94 & \greenc 91.48 & 94.33 & \greenc 87.56 & 92.13 & \greenc 94.87 & 93.18 \down{1.08} & \greenc 92.56 \down{0.95}\\
\ourmethod (PR=0.5\%) & 94.36 & \grayc 92.91 & 93.32 & \greenc 90.96 & 93.87 & \greenc 84.04 & 94.52 & \greenc 86.82 & 91.99 & \greenc 84.63 & 93.43 \down{0.93} & \greenc 86.61 \down{6.30}\\
\bottomrule
\end{tabular}
\end{table*}

\begin{table*}[htb]
\centering
\caption{Fine-tuning-based mitigations against backdoored ResNet18 on CIFAR10.}
\label{tab:fine_tuning}
\renewcommand{\arraystretch}{0.95}
\begin{tabular}{ccccccccccccc}
\toprule
\multirow{2}{*}{Attack} & \multicolumn{2}{c}{vanilla FT} & \multicolumn{2}{c}{FT-SAM~\cite{Zhu2023ftsam}} & \multicolumn{2}{c}{I-BAU~\cite{zeng2022ibau}} & \multicolumn{2}{c}{FST~\cite{min2023fst}} & \multicolumn{2}{c}{BTI-DBF(U)~\cite{xu2024btidbf}} & \multicolumn{2}{c}{Average}\\
\cmidrule(lr){2-3}
\cmidrule(lr){4-5}
\cmidrule(lr){6-7}
\cmidrule(lr){8-9}
\cmidrule(lr){10-11}
\cmidrule(lr){12-13}
~ & BA & ASR & BA & ASR & BA & ASR & BA & ASR & BA & ASR & BA & ASR \\
\midrule
BadNets~\cite{Gu_2019_badnet} & 91.07 & \pinkc 43.96 & 92.01 & \pinkc 2.84 & 92.60 & \greenc 76.02 & 92.40 & \pinkc 13.10 & 91.26 & \pinkc 13.12 & 91.87 \down{1.26} & \pinkc  29.81 \down{70.19}\\
Blend~\cite{chen_2017_blend} & 91.64 & \greenc 99.61 & 92.52 & \pinkc 1.73 & 91.84 & \pinkc 8.84 & 93.40 & \greenc 100 & 91.86 & \greenc 100 & 92.25 \down{2.17} & \greenc 62.04 \down{37.96}\\
WaNet~\cite{nguyen2021wanet} & 91.11 & \pinkc 0.99 & 90.89 & \pinkc 1.03 & 87.98 & \pinkc 0.81 & 92.17 & \pinkc 0.04 & 90.30 & \pinkc 4.89 & 90.49 \down{3.11} & \pinkc 1.55 \down{97.82}\\
IAD~\cite{nguyen2020inputaware} & 90.83 & \pinkc 2.16 & 92.18 & \pinkc 2.87 & 88.4 & \pinkc 15.68 & 91.29 & \pinkc 0.00 & 89.54 & \pinkc 1.59 & 90.45 \down{2.43} & \pinkc 4.46 \down{92.64}\\
AdvDoor~\cite{zhang2021advdoor} & 91.25 & \greenc 68.68 & 92.18 & \pinkc 1.23 & 89.29 & \pinkc 16.99 & 91.06 & \greenc 99.99 & 90.25 & \greenc 100 & 90.81 \down{3.16} & \pinkc 57.38 \down{42.62}\\
Bpp~\cite{Wang_2022_bpp} & 91.36 & \pinkc 3.40 & 91.38 & \pinkc 1.00 & 92.06 & \pinkc 6.46 & 93.23 & \pinkc 26.83 & 90.61 & \pinkc 2.73  & 91.73 \down{2.46} & \pinkc 8.08 \down{91.85}\\
LC~\cite{turner2019lc} & 90.26 & \greenc 88.52 & 91.46 & \pinkc 1.91 & 85.87 & \pinkc 5.11 & 91.80 & \pinkc 13.11 & 90.71 & \pinkc 4.37 & 90.02 \down{4.29} & \pinkc 22.60 \down{77.40}\\
Narcissus~\cite{zeng2023narcisus} & 91.70 & \greenc 92.91 & 91.76 & \pinkc 23.98 & 91.48 & \pinkc 51.74 & 90.06 & \pinkc 54.22 & 90.94 & \greenc 98.11 & 91.19 \down{2.39} & \greenc 64.19 \down{35.45}\\
Adap-blend~\cite{qi2022revisitingadaptive} & 92.42 & \greenc 98.73 & 91.23 & \pinkc 22.4 & 85.38 & \pinkc 37.31 & 90.91 & \pinkc 1.19 & 89.17 & \pinkc 7.09 & 89.82 \down{2.92} & \pinkc 33.34 \down{66.33}\\
SSDT~\cite{mo2024ssdt} & 93.74 & \pinkc 0.70 & 93.15 & \pinkc 0.60 & 90.27 & \pinkc 3.10 & 92.85 & \pinkc 0.20 & 90.79 & \pinkc 1.40 & 92.16 \down{1.54} & \pinkc 1.20 \down{89.10}\\
DFST~\cite{Cheng2021dfst} & 95.01 & \pinkc 2.07 & 94.70 & \pinkc 0.00 & 89.75 & \pinkc 19.11 & 93.06 & \pinkc 2.66 & 90.41 & \pinkc 22.34 & 92.59 \down{2.64} & \pinkc 9.24 \down{90.76}\\
DFBA~\cite{cao2024dfba} & 86.68 & \pinkc 10.10 & 86.03 & \pinkc 5.24 & 85.48 & \greenc 100 & 82.76 & \pinkc 57.62 & 84.49 & \greenc 100 & 85.09 \down{3.90} & \pinkc 54.59 \down{45.41} \\
\midrule
\ourmethod (PR=5\%) & 91.75 & \greenc 94.28 & 92.02 & \greenc 80.07 & 90.39 & \greenc 93.92 & 93.27 & \greenc 99.92 & 91.88 & \greenc 99.00 & 91.86 \down{1.57} & \greenc 93.44 \down{4.60}\\
\ourmethod (PR=1\%) & 91.41 & \greenc 85.52 & 92.83 & \greenc 79.17 & 87.89 & \greenc 91.34 & 93.21 & \greenc 96.59 & 90.66 & \greenc 88.69 & 91.20 \down{3.06} & \greenc 88.26 \down{5.25}\\
\ourmethod (PR=0.5\%) & 91.42 & \greenc 82.96 & 92.34 & \greenc 76.92 & 89.83 & \greenc 79.68 & 93.44 & \greenc 92.71 & 90.39 & \greenc 91.83 & 91.48 \down{2.88} & \greenc 84.82 \down{8.09} \\
\bottomrule
\end{tabular}
\end{table*}

\begin{table*}[htb]
\centering
\caption{Backdoor performance of \ourmethod and baseline attacks on ImageNet200 and GTSRB.}
\label{tab:datasets_allattacks}
\renewcommand{\arraystretch}{0.95}
\begin{tabular}{cccccccccccc}
\toprule
\multirow{2}{*}{Datasets} & \multirow{2}{*}{Attack} & \multicolumn{2}{c}{No Defense} & \multicolumn{2}{c}{FT-SAM~\cite{Zhu2023ftsam}} & \multicolumn{2}{c}{I-BAU~\cite{zeng2022ibau}} & \multicolumn{2}{c}{CLP~\cite{zheng2022clp}} & \multicolumn{2}{c}{Average}\\
\cmidrule(lr){3-4}
\cmidrule(lr){5-6}
\cmidrule(lr){7-8}
\cmidrule(lr){9-10}
\cmidrule(lr){11-12}
~ & ~ & BA & ASR & BA & ASR & BA & ASR & BA & ASR & BA & ASR\\
\midrule
\multirow{9}{*}{ImageNet200} & BadNets~\cite{Gu_2019_badnet} & 80.65 & \grayc 91.03 & 79.89 & \pinkc 2.21 & 70.28 & \pinkc 26.06 & 70.74 & \greenc 64.86 & 73.64 \down{7.01} & \pinkc 31.04 \down{59.99}\\
~ & Blend~\cite{chen_2017_blend} & 80.70 & \grayc 95.63 & 80.19 & \pinkc 0.39 & 76.13 & \pinkc 30.81 & 80.02 & \pinkc 23.38 & 78.78 \down{1.92} & \pinkc 18.19 \down{77.44}\\
~ & WaNet~\cite{nguyen2021wanet} & 81.24 & \grayc 99.97 & 80.41 & \pinkc 0.66 & 75.67 & \pinkc 47.27 & 77.18 & \greenc 99.78 & 77.75 \down{3.49} & \pinkc 49.24 \down{50.73}\\
~ & IAD~\cite{nguyen2020inputaware} & 79.74 & \grayc 99.98 & 75.49 & \pinkc 0.68 & 77.44 & \pinkc 15.18 & 76.97 & \greenc 84.49 & 76.63 \down{3.11} & \pinkc 33.45 \down{66.53}\\
~ & AdvDoor~\cite{zhang2021advdoor} & 80.72 & \grayc 100 & 79.52 & \greenc 98.90 & 74.03 & \greenc 61.31 & 77.90 & \greenc 100 & 77.15 \down{3.57} & \greenc 86.74 \down{13.26}\\
~ & Bpp~\cite{Wang_2022_bpp} & 81.36 & \grayc 92.74 & 79.37 & \pinkc 1.05 & 76.53 & \pinkc 3.21 & 80.10 & \pinkc 2.34 & 78.67 \down{2.69} & \pinkc 2.19 \down{90.55}\\
~ & Narcissus~\cite{zeng2023narcisus} & 81.73 & \grayc 81.28 & 80.00 & \greenc 83.37 & 77.03 & \pinkc 56.19 & 80.99 & \greenc 86.37 & 79.34 \down{2.39} & \greenc 75.31 \down{5.97}\\
~ & SSDT~\cite{mo2024ssdt} & 75.45 & \grayc 100 & 78.19 & \greenc 76.00 & 76.26 & \pinkc 22.00 & 76.02 & \greenc 94.00 & 76.82 \up{1.37} & \greenc 64.00 \down{36.00}\\
\cmidrule(ll){2-12}
~ & \ourmethod & 80.92 & \grayc 94.11 & 79.05 & \greenc 95.05 & 76.89 & \greenc 87.75 & 80.29 & \greenc 93.83 & 78.74 \down{2.18} & \greenc 92.21 \down{1.9}\\
\midrule
\multirow{9}{*}{GTSRB} & BadNets~\cite{Gu_2019_badnet} & 97.19 & \grayc 100 & 95.57 & \pinkc 0.48 & 92.02 & \pinkc 29.22 & 96.38 & \pinkc 0.47 & 94.66 \down{2.53} & \pinkc 10.06 \down{89.94}\\
~ & Blend~\cite{chen_2017_blend} & 95.92 & \grayc 100 & 93.36 & \pinkc 0.21 & 92.64 & \pinkc 38.27 & 93.21 & \pinkc 0.00 & 93.07 \down{2.85} & \pinkc 12.83 \down{87.17}\\
~ & WaNet~\cite{nguyen2021wanet} & 98.69 & \grayc 99.77 & 92.18 & \pinkc 0.45 & 91.25 & \pinkc 0.00 & 90.14 & \pinkc 18.14 & 91.19 \down{7.50} & \pinkc 6.19 \down{93.58}\\
~ & IAD~\cite{nguyen2020inputaware} & 99.08 & \grayc 99.65 & 92.72 & \pinkc 0.10 & 90.11 & \pinkc 0.35 & 98.08 & \pinkc 14.63 & 93.64 \down{5.44} & \pinkc 5.03 \down{94.62}\\
~ & AdvDoor~\cite{zhang2021advdoor} & 95.80 &\grayc  99.99 & 93.94 & \pinkc 32.26 & 92.67 & \pinkc 38.20 & 90.09 & \greenc 66.39 & 92.23 \down{3.57} & \pinkc 45.62 \down{54.37}\\
~ & Bpp~\cite{Wang_2022_bpp} & 98.69 & \grayc 99.93 & 91.27 & \pinkc 0.00 & 92.61 & \pinkc 0.23 & 97.16 & \pinkc 2.29 & 93.68 \down{5.01} & \pinkc 0.84 \down{99.09}\\
~ & Narcissus~\cite{zeng2023narcisus} & 95.60 & \grayc 97.18 & 93.61 & \pinkc 54.55 & 92.87 & \greenc 80.74 & 93.99 & \greenc 97.60 & 93.49 \down{2.11} & \greenc 77.63 \down{19.55}\\
~ & SSDT~\cite{mo2024ssdt} & 96.02 & \grayc 77.78 & 93.11 &\pinkc 0.00 & 90.82 & \pinkc 0.00 & 94.65 & \pinkc 19.31 & 92.86 \down{3.16} & \pinkc 6.44 \down{71.34}\\
\cmidrule(ll){2-12}
~ & \ourmethod & 95.83 & \grayc 95.36 & 93.80 & \greenc 71.84 & 93.13 & \greenc 94.30 & 91.28 & \greenc 93.19 & 92.74 \down{3.09} & \greenc 86.44 \down{8.92}\\
\bottomrule
\end{tabular}
\end{table*}

\section{Experimental Evaluation}
\label{sec:experiments}

\subsection{Experimental Setup}
\label{sec:exp_setup}

\noindent
\textbf{Datasets and Architectures.}
We follow the common settings in existing backdoor attacks and defenses and conduct experiments on CIFAR10~\cite{krizhevsky2009cifar10}, GTSRB~\cite{Stallkamp2012gtsrb}, and a subset of ImageNet~\cite{jia2009imagenet} with 200 classes and 1,300 images per class (ImageNet200). 
More details about the datasets can be found in \appcite{appendix:datasets}.
The primary evaluation is performed using ResNet18~\cite{He_2016_resnet}.
Moreover, we evaluate \ourmethod using four additional architectures, VGG16~\cite{Karen_2015_vgg}, DenseNet121~\cite{huang2017densely}, EfficientNet-B0~\cite{tan2019efficientnet}, and one recent architecture InceptionNeXt~\cite{Yu2024inceptionnext} (see Table~\ref{tab:architectures} in \appcite{sec:diff_arch_surrogate}).
We also evaluate \ourmethod with large and transformer-based models (see Table~\ref{tab:largearch} in \appcite{sec:large_vit}).

\noindent
\textbf{Attack Baselines.}
\ourmethod is compared with 12 representative attacks: BadNets~\cite{Gu_2019_badnet}, Blend~\cite{chen_2017_blend}, WaNet~\cite{nguyen2021wanet}, IAD~\cite{nguyen2020inputaware}, AdvDoor~\cite{zhang2021advdoor}, BppAttack~\cite{Wang_2022_bpp}, LC~\cite{turner2019lc}, Narcissus~\cite{zeng2023narcisus}, Adap-Blend~\cite{qi2022revisitingadaptive}, SSDT~\cite{mo2024ssdt}, DFST~\cite{Cheng2021dfst}, and DFBA~\cite{cao2024dfba}.
The default poisoning rate is set at 5\% (of the training set) for all attacks following previous work~\cite{xu2024btidbf,xu2024ban}.
Additionally, \ourmethod is evaluated under various poisoning rates to provide a thorough analysis of its effectiveness.
Following related works, the training schedule for attacks is 200 epochs when using CIFAR10 and GTSRB, and 100 epochs for ImageNet200. 
We use 1,000 images as the validation set to select the best-performing checkpoint.
More implementation details are provided in \appcite{appendix:back_attacks}.

\noindent
\textbf{Defense Baselines.}
We evaluate \ourmethod and baseline attacks with 17 defenses, including \textbf{four pruning-based} methods (FP~\cite{liu2018finepruning}, ANP~\cite{Wu_2021_anp}, CLP~\cite{zheng2022clp}, and RNP~\cite{li2023rnp}),
\textbf{five fine-tuning-based} methods (vanilla FT, FT-SAM~\cite{Zhu2023ftsam}, I-BAU~\cite{zeng2022ibau}, FST~\cite{min2023fst}, and BTI-DBF (U)~\cite{xu2024btidbf}), \textbf{five backdoor model detections} (NC~\cite{wang_2019_nc}, Tabor~\cite{guo_2020_tabor}, FeatureRE~\cite{wang2022featurere}, Unicorn~\cite{wang2023unicorn}, and BTI-DBF~\cite{xu2024btidbf}), \textbf{two backdoor input detections} (Scale-up~\cite{guo2023scaleup} and IBD-PSC~\cite{hou2024ibdpsc}), and a \textbf{proactive defense} CT~\cite{qi2023proactivedetection}.
Following their default settings, BTI-DBF~\cite{xu2024btidbf} and FP~\cite{liu2018finepruning} use 5\% of training data, and other defenses use 1\% of training data for detection or mitigation.
CLP is a data-free backdoor pruning tool that uses no clean data.
CT has access to the complete training set without knowing which samples are poisoned and can also interact with the model during training.
Backdoor defense details and hyperparameters can be found in \appcite{appendix:defenses}.

\subsection{Main Results on Backdoor Mitigation}
\label{sec:main_results}

All evaluated backdoor attacks are ineffective against at least one parameter-space backdoor defense on the CIFAR10, as demonstrated in Tables~\ref{tab:pruning} and~\ref{tab:fine_tuning}. 
It suggests that common backdoor attacks designed to be stealthy in input and feature spaces are vulnerable to parameter-space defenses. 
Given that all backdoor behaviors are embedded in parameters of backdoored models, this finding suggests that future backdoor attacks should consider parameter-space defenses as a standard step to evaluate comprehensive stealthiness. 

Not surprisingly, \ourmethod performs better than all baseline attacks when considering evaluated backdoor defenses since \ourmethod is designed to consider comprehensive stealthiness. 
On four pruning-based mitigations, \ourmethod achieves 7.18\% higher ASR on average than the best backdoor attack, Narcissus. 
On five fine-tuning mitigations that show more powerful defense capability than pruning-based mitigations, \ourmethod achieves 29.25\% higher ASR on average than Narcissus.
In addition, \ourmethod bypasses the five model detection and two input-space detections (see Section~\ref{sec:detection}).

\noindent \textbf{Pruning-based mitigation.}
We take a closer look at the details of pruning-based backdoor mitigation experiments in Table~\ref{tab:pruning}, presenting the results of all attacks against four pruning-based defenses.
BadNets and Blend perform better on average than input-space stealthy attacks, e.g., WaNet and Bpp, because input-space stealthy attacks introduce significant separability in the feature space (see Figures~\ref{fig:feature_tsne} and~\ref{fig:featureloss}).
Across all pruning-based defenses, FP performs the worst, as expected, since it follows regular model pruning practice and is not a tailored backdoor pruning method.

\begin{figure}[t]
\centering
\includegraphics[width = 1\linewidth]{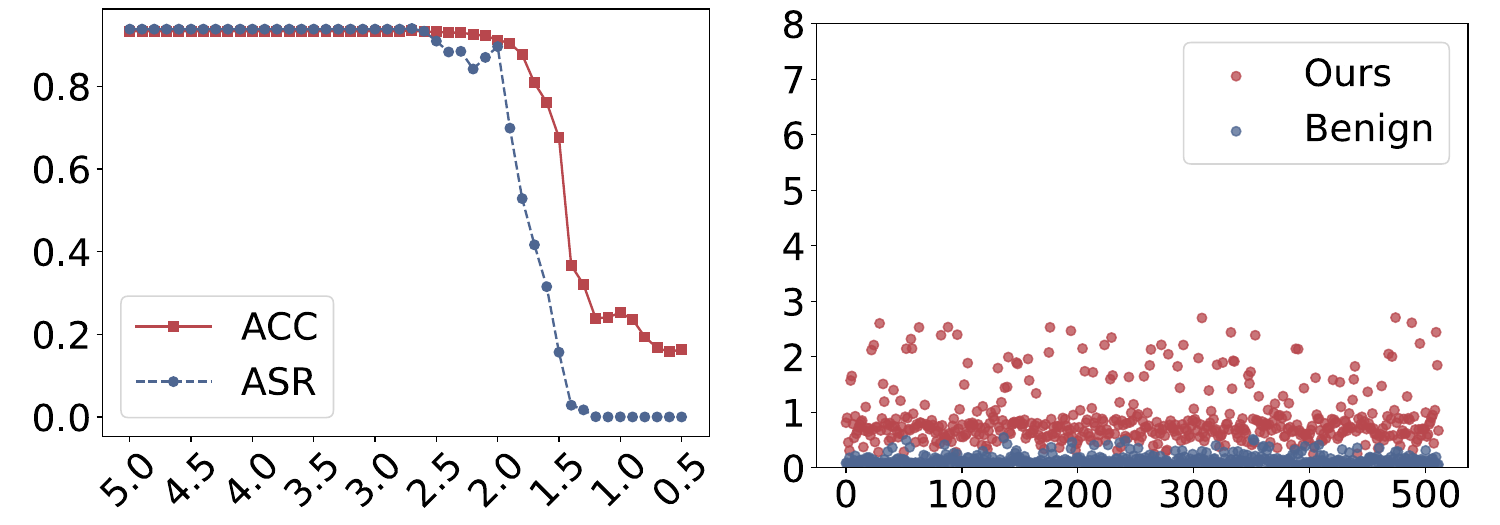}
\caption{The left is the performance of pruning neurons with high TAC values using different thresholds for \ourmethod. The right is the TAC analysis of \ourmethod on 512 neurons.}
\label{fig:tac_plot}
\end{figure}

\noindent
\textbf{Fine-tuning-based mitigations.}
Table~\ref{tab:fine_tuning} presents the backdoor performance against five fine-tuning-based defenses.
In general, fine-tuning-based defenses are more effective than pruning-based defenses.
For example, Narcissus and Adap-Blend can achieve ASRs higher than 60\% against three out of four pruning-based defenses but are much less effective against most fine-tuning-based methods.
FT-SAM is the most effective across all defenses, as shown in Tables~\ref{tab:pruning} and~\ref{tab:fine_tuning}, being able to compromise the effectiveness of all attack baselines.
One important reason is that FT-SAM adopts Sharpness-Aware Minimization~\cite{foret2021sam} to adjust the outlier of weight norm (large norms) to remove the potential backdoor.
Larger weights of neurons are introduced by existing attacks to guarantee a high ASR~\cite{liu_2019_abs}, which also causes large differences when receiving benign and backdoor inputs (see Figure~\ref{fig:tac_plot}).
\ourmethod can bypass FT-SAM, as expected, since it deliberately decreases the weights of backdoor neurons, compromising the core working mechanism of FT-SAM.

\begin{figure*}[t]
\centering
\includegraphics[width = 1\linewidth]{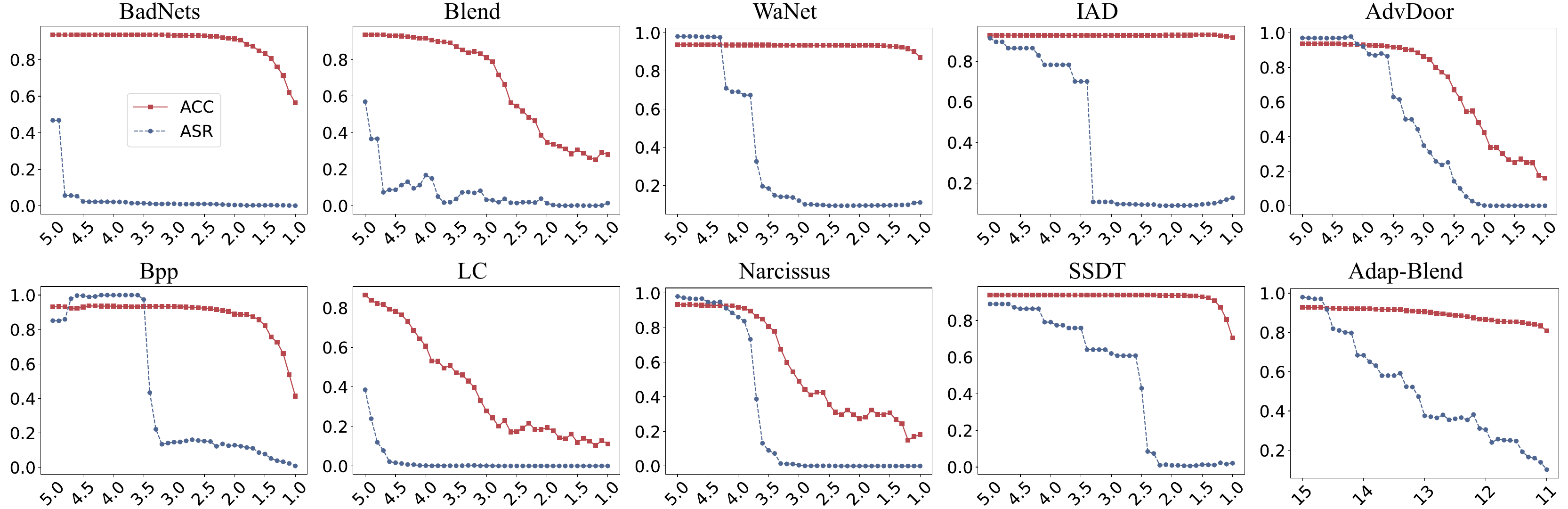}
\caption{Pruning neurons with high TAC values using different thresholds (the $x$ axis).}
\label{fig:tac_prune_thresholds}
\end{figure*}

\noindent
\textbf{Comparison with Supply-Chain Attacks.}
Sharing a similar threat model to supply-chain attacks, we compare \ourmethod and three state-of-the-art supply-chain attacks (SSDT~\cite{mo2024ssdt}, DFST~\cite{Cheng2021dfst}, and DFBA~\cite{cao2024dfba}), where these attacks are also designed to be robust against backdoor defenses.
In particular, DFST~\cite{Cheng2021dfst} proposes to include a controlled detoxification technique in the training process, which restrains the model from picking up simple features.
DFBA~\cite{cao2024dfba} directly modifies a few parameters of a classifier to inject a backdoor.
SSDT~\cite{mo2024ssdt} introduces additional terms in the loss for the Source-Specific and Dynamic-Triggers (i.e., SSDT) attack, which obscures the difference between normal samples and malicious samples.
Tables~\ref{tab:pruning} and~\ref{tab:fine_tuning} also include the performance of supply-chain attacks against pruning- and fine-tuning-based defenses.
It is clear that existing backdoor defenses can defeat supply-chain attacks.
The ASR of DFST~\cite{Cheng2021dfst}, DFBA~\cite{cao2024dfba}, and SSDT~\cite{mo2024ssdt} are decreased to less than 10\% while the BA drop is less than 3\%.

\noindent
\textbf{On ImageNet200 and GTSRB.}
Real-world classification tasks may involve more categories, such as GTSRB (43 classes) and ImageNet200 (200 classes), and the percentage of each class in the dataset will commonly be much less than 10\%.
We target InceptionNext-Small on Imagenet200 and ResNet18 on GTSRB.
The $l_\infty$ norm perturbation budget of UPGD is $\epsilon=16$ for GTSRB and $\epsilon=8$ for ImageNet200 to achieve imperceptible perturbations.
Table~\ref{tab:datasets_allattacks} demonstrates that \ourmethod is still effective on datasets with more classes and higher resolutions, especially against the most powerful parameter-space defense, FT-SAM.

\subsection{Adaptive Defenses} 

As \ourmethod includes UPGD trigger and Adversarial Backdoor Injection (ABI), we consider two adaptive defenses targeting those two components.

\noindent
\textbf{TAC pruning targeting the UPGD trigger.}
First, we use the UPGD trigger information to build a new pruning method based on the TAC values.
The TAC values are calculated using the backdoor trigger, making the TAC highly adaptive to evaluating any backdoor attack. 
In particular, we prune neurons with high TAC values in the backdoored model, i.e., removing the neurons more sensitive to the UPGD trigger.
Figure~\ref{fig:tac_plot} shows the pruning results of \ourmethod.
The left figure provides the pruning results.
The right figure contains the TAC values plots of neurons in the $4_{th}$ layer (the layer before the classification head) of ResNet18.
We show that pruning neurons with high TAC values decreases benign accuracy, which means the backdoor neurons are not easily distinguishable from benign neurons without harming benign performance. 
The analysis supports our statement that \ourmethod spreads the backdoor to more neurons instead of a few prominent ones. 
In \appcite{appendix:tac_others}, we provide sorted TAC value plots in Figure~\ref{fig:tac_sort}, showing that prominent neurons with high TAC values are rather limited in \ourmethod.

\noindent
\textbf{Neuron noise targeting ABI.}
Second, we consider adding noise to neurons of \ourmethod models as an adaptive defense, as the ABI component involves operations on neuron weights.
Specifically, we add noise to the weights and biases of batch normalization layers.
The range of noise is limited by $[-\epsilon_{noise},\epsilon_{noise}]$.
Table~\ref{tab:noise_adaptive} provides the performance of \ourmethod and the benign model under different noise levels.
With increased noise, the benign accuracy of both benign and \ourmethod models is decreased, but the ASR of \ourmethod remains high.
This result also supports the stealthiness of \ourmethod in the parameter space.

\subsection{Backdoor Analysis}
\label{sec:backdoor_analysis}

This section analyzes why baseline attacks are ineffective against parameter space backdoor defenses, and why \ourmethod performs better, according to the TAC values and the weights' change after applying backdoor defenses.

\begin{table}[tb]
\centering
\caption{Adaptive defense against \ourmethod using noise on neurons.}
\label{tab:noise_adaptive}
\begin{tabular}{ccccccccccccccccc}
\toprule
\multirow{2}{*}{\tabincell{c}{Method}} &  \multicolumn{2}{c}{Benign} & \multicolumn{2}{c}{\ourmethod}\\
\cmidrule(lr){2-3}
\cmidrule(lr){4-5}
 ~ & BA & ASR & BA & ASR\\
  \midrule
 \multicolumn{1}{c}{$\epsilon_{noise}=0.0$} & 94.76 & - & 94.16 & 98.04 \\
\multicolumn{1}{c}{$\epsilon_{noise}=0.1$} & 94.54 & - & 93.88 & 97.99 \\
\multicolumn{1}{c}{$\epsilon_{noise}=0.2$} & 93.82 & - & 93.18 & 96.22 \\
\multicolumn{1}{c}{$\epsilon_{noise}=0.3$} & 91.21 & - & 90.05 & 99.08\\
\multicolumn{1}{c}{$\epsilon_{noise}=0.4$} & 85.65 & - & 88.33 & 99.42\\
\multicolumn{1}{c}{$\epsilon_{noise}=0.5$} & 82.43 & - & 84.30 & 99.74 \\
\bottomrule
\end{tabular}
\end{table}

\noindent
\textbf{Pruning neurons with prominent TAC values for baseline attacks.}
Section~\ref{sec:lack} demonstrates the existence of prominent neurons with high TAC values in backdoor models.
Note that TAC represents the strongest type of backdoor defense, where the exact trigger information is exploited. 
In this section, we show that pruning prominent neurons could mitigate all baseline attacks. 
Specifically, we assign zero to the neuron's weight if its TAC value exceeds a certain threshold.
Figure~\ref{fig:tac_prune_thresholds} shows the pruning performance using different thresholds for ten baseline attacks.
We can observe that for all ten baseline attacks, one threshold can always be found where the ACC is high and the ASR is low. 
It indicates that the backdoor effect of the ten attacks can be erased while maintaining good performance on benign samples if these prominent neurons are pruned.
Thus, we confirm the existence of prominent neurons, which corresponds to the backdoor effect.
More importantly, these backdoor neurons can be disentangled from benign neurons, so these baseline attacks are not stealthy concerning neuron weights, i.e., not stealthy in the parameter space.

\begin{figure}[t]
\centering
\includegraphics[width = 1\linewidth]{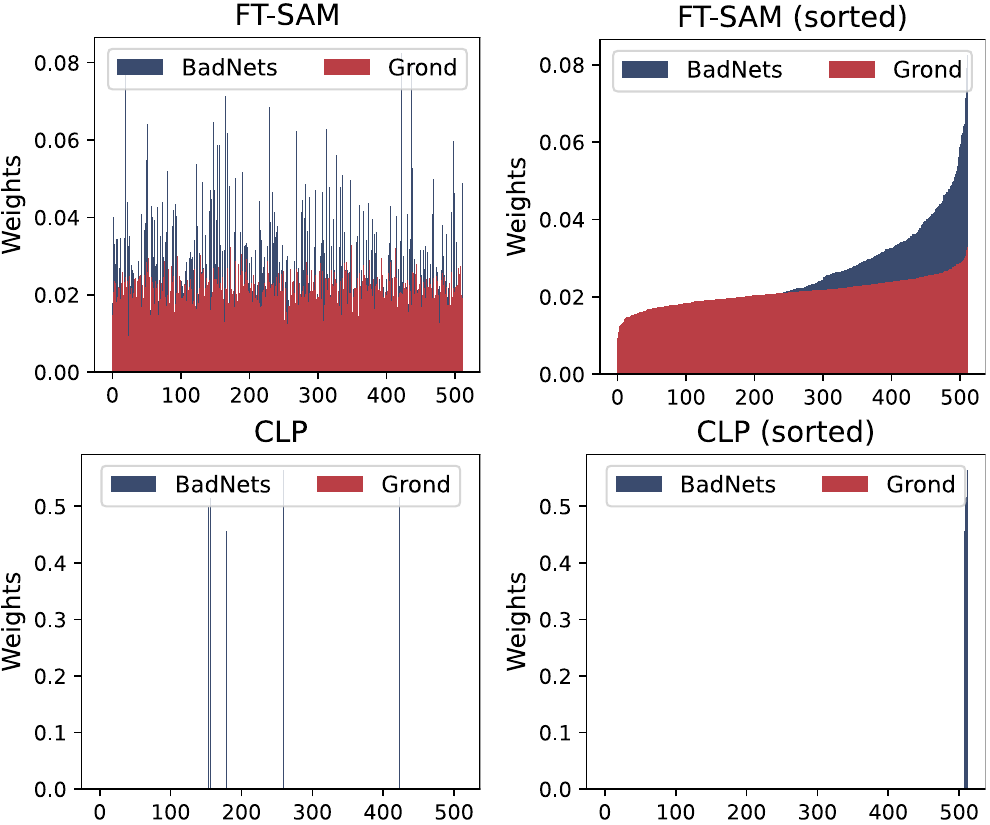}
\caption{The weight changes after backdoor defenses. More results with other attacks in Figure~\ref{fig:weight_changes_all} in \appcite{sec:weight_changes_all}.}
\label{fig:weight_changes}
\end{figure}

\noindent
\textbf{\ourmethod's weights are more difficult to be modified by backdoor defense.}
This section analyzes the changes while applying pruning- and fine-tuning-based defenses to baseline attacks and \ourmethod.
Our goal is to demonstrate that the baseline attacks can be significantly affected by defenses, but \ourmethod can resist these defenses.
In Figure~\ref{fig:weight_changes}, we record the changes in the weights of 512 neurons (in layer 4 of ResNet18) after applying two types of defenses, FT-SAM~\cite{Zhu2023ftsam} and CLP~\cite{zheng2022clp}, as they both focus on the backdoor-related neurons.
In contrast to BadNets, the weight changes in \ourmethod model after FT-SAM fine-tuning are smaller.
In the sorted changes, it is clear that there are a few neurons for BadNets that correspond to significant changes, which is not the case for \ourmethod.
In the second row of Figure~\ref{fig:weight_changes}, the CLP can find a few neurons relevant to backdoor but cannot find these for \ourmethod.
In addition, pruning-based methods (as shown by CLP results in Figure~\ref{fig:weight_changes}) only improve a few neurons, while fine-tuning methods can update all neurons.
We conjecture this is why fine-tuning-based defenses perform better than pruning-based defenses in Tables~\ref{tab:pruning} and~\ref{tab:fine_tuning}. 
More results with other baseline attacks are provided in Figure~\ref{fig:weight_changes_all} in \appcite{sec:weight_changes_all}.

\begin{table}[t]
\centering
\caption{Backdoor detection performance on CIFAR10. 20 ResNet18 models are trained at each poisoning rate. Bd. refers to the number of models determined as backdoor models. Acc. refers to the detection accuracy.}
\label{tab:detection}
\begin{tabular}{ccccccccccc}
\toprule
\multirow{2}{*}{Defense} & \multicolumn{2}{c}{PR=5\%} & \multicolumn{2}{c}{PR=1\%} & \multicolumn{2}{c}{PR=0.5\%}\\
\cmidrule(lr){2-3}
\cmidrule(lr){4-5}
\cmidrule(lr){6-7}
~ & Bd. & Acc. & Bd. & Acc. & Bd. & Acc.\\
\midrule
NC~\cite{wang_2019_nc} & 5 & 25\% & 2 & 10\% & 1 & 5\%\\
Tabor~\cite{guo_2020_tabor} & 5 & 25\% & 2 & 10\% & 0 & 0\%\\
FeatureRE~\cite{wang2022featurere} & 0 & 0\% & 0 & 0\% & 0 & 0\%\\
Unicorn~\cite{wang2023unicorn} & 0 & 0\% & 0 & 0\% & 0 & 0\%\\
BTI-DBF~\cite{xu2024btidbf} & 3 & 15\% & 5 & 25\% & 3 & 15\%\\
\bottomrule
\end{tabular}
\end{table}

\subsection{ABI Improves Common Backdoor Attacks}
\label{sec:abi_improves}

\begin{figure*}[htb]
\centering
\includegraphics[width = 1\linewidth]{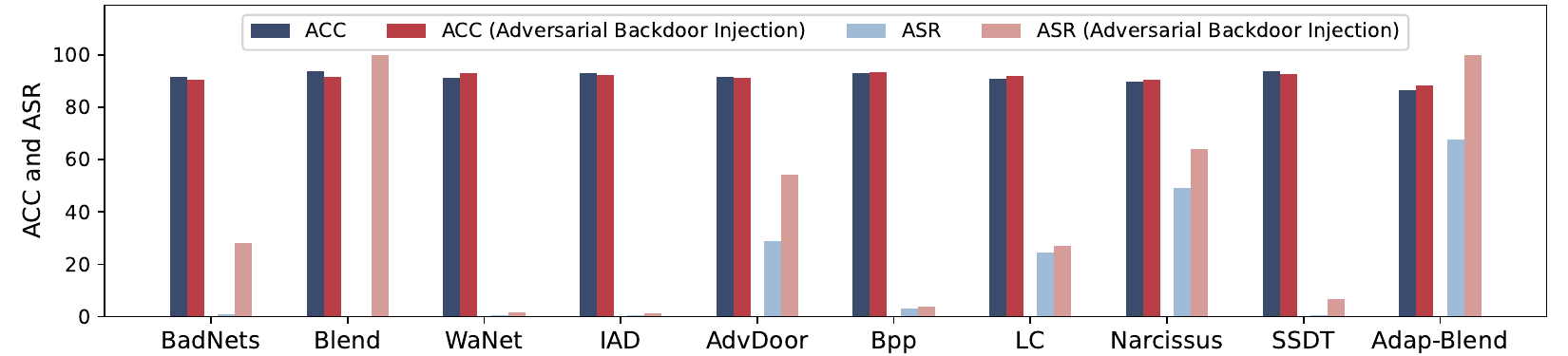}
\caption{BA and ASR of backdoor attacks before and after ABI against parameter-space defense ANP.}
\label{fig:clp_otherattacks}
\end{figure*}

In this section, we show that our Adversarial Backdoor Injection (ABI) strategy generalizes to all evaluated common backdoor attacks. 
We combine the \abi module with baseline attacks to improve their resistance against parameter-space defenses.
Figure~\ref{fig:clp_otherattacks} demonstrates that \abi is effective for all attacks when evaluating against the parameter-space defense ANP, where ASRs increase after adversarial injection, especially for BadNets, Blend, AdvDoor, Narcissus, and Adap-Blend.
The improvement for feature space attacks (WaNet, IAD, and Bpp) is incremental. 
We speculate that feature space attacks rely too much on prominent features, as their modification in the input space is minor.
To activate the backdoor with such minor input modifications, the prominent features are required in the feature space.
In addition, Figure~\ref{fig:clp_otherattacks_nodefense} in \appcite{appendix:advinjection_otherattack} shows the results of \abi without defense, demonstrating that it does not harm in general the BA and ASR when no defense is applied.
Following our findings, we suggest that future backdoor attacks can use \abi to increase parameter-space stealthiness.

\subsection{Backdoor Detection}
\label{sec:detection}

Following previous works~\cite{xu2024btidbf,xu2024ban}, we choose five representative backdoor model detections for evaluation.
We use 20 models for each poisoning rate with different random seeds.
Then, we report the number of models detected as backdoor models out of the 20.
Table~\ref{tab:detection} shows that all detections fall short when detecting \ourmethod.
In particular, NC~\cite{wang_2019_nc}, Tabor~\cite{wang_2019_nc}, and BTI-DBF~\cite{xu2024btidbf} can detect a small part of backdoored models, while FeatureRE~\cite{wang2022featurere} and Unicorn cannot detect any of them.
For featureRE~\cite{wang2022featurere}, we conjecture it is over-dependent on the separability in the feature space, but \ourmethod does not rely on prominent backdoor features according to Figure~\ref{fig:featureloss} in the Appendix~\appcite{appendix:featureloss}.
For Unicorn~\cite{wang2023unicorn}, the false positive rate is high, and it tends to report every class as the backdoor target, even on models trained with benign data only.
Except for model detection, \ourmethod can also bypass input-space detections as demonstrated in \appcite{sec:detect_input}.

\begin{table}[tb]
\centering
\caption{Comparison with different strategies for the generation of backdoor triggers.}
\label{tab:diff_trigger}
\begin{tabular}{cccccccccc}
\toprule
\multirow{2}{*}{Strategy} & \multicolumn{2}{c}{No Defense} & \multicolumn{2}{c}{CLP~\cite{zheng2022clp}} & \multicolumn{2}{c}{FT-SAM~\cite{Zhu2023ftsam}}\\
\cmidrule(lr){2-3}
\cmidrule(lr){4-5}
\cmidrule(lr){6-7}
 ~ & BA & ASR & BA & ASR & BA & ASR\\
 \midrule
 Random noise & 94.24 & 1.28 & 94.13 & 0.97 & 93.90 & 1.84\\
 PGD  & 94.77 & 69.33 & 92.57 & 46.63 & 92.40 & 24.56\\
 UPGD & 93.43 & 98.04 & 93.29 & 87.89 & 92.02 & 80.07\\
\bottomrule
\end{tabular}
\end{table}

\begin{table}[tb]
\centering
\setlength\tabcolsep{2.5pt}
\caption{The semantic trigger (an automobile image) with different strategies for \ourmethod. The ``Target Same'' refers to the target class being the same as the semantic trigger, i.e., automobile. The ``Target Diff.'' refers to the target class being different (i.e., airplane) from the semantic trigger.}
\label{tab:semantic_trigger}
\begin{tabular}{clccccccccc}
\toprule
\multicolumn{2}{c}{\multirow{2}{*}{Strategies for \begin{minipage}[b]{0.06\columnwidth}
		\centering
		\raisebox{-.25\height}{\includegraphics[width=\linewidth]{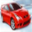}}
	\end{minipage} }} & \multicolumn{2}{c}{No Defense} & \multicolumn{2}{c}{CLP~\cite{zheng2022clp}} & \multicolumn{2}{c}{FT-SAM~\cite{Zhu2023ftsam}}\\
\cmidrule(lr){3-4}
\cmidrule(lr){5-6}
\cmidrule(lr){7-8}
 ~ & ~ & BA & ASR & BA & ASR & BA & ASR\\
\midrule
\multirow{2}{*}{Clean-Label} & Target Same & 94.20 & 74.64 & 93.97 & 70.66 & 91.68 & 9.38\\
~ & Target Diff. & 93.82 & 94.42 & 93.75 & 93.50 & 90.92 & 31.37\\
\midrule
\multirow{2}{*}{Dirty-Label} & Target Same & 93.77 & 100 & 93.88 & 100 & 91.40 & 12.17\\
~ & Target Diff. & 94.25 & 100 & 94.14 & 100 & 91.34 & 7.39\\
\bottomrule
\end{tabular}
\end{table}

\begin{table}[tb]
\centering
\setlength\tabcolsep{3pt}
\caption{Ablation study for \ourmethod.}
\label{tab:ablation}
\begin{tabular}{clccccccccc}
\toprule
\multirow{2}{*}{Arch} & \multirow{2}{*}{Method} & \multicolumn{2}{c}{No Defense} & \multicolumn{2}{c}{CLP~\cite{zheng2022clp}} & \multicolumn{2}{c}{FT-SAM~\cite{Zhu2023ftsam}}\\
\cmidrule(lr){3-4}
\cmidrule(lr){5-6}
\cmidrule(lr){7-8}
 ~ & ~ & BA & ASR & BA & ASR & BA & ASR\\
\midrule
\multirow{2}{*}{ResNet18} & UPGD & 93.86 & 98.61 & 91.15 & 3.97 & 91.80 & 51.77\\
~ & +ABI & 93.43 & 98.04 & 93.29 & 87.89 & 92.02 & 80.07\\
\midrule
\multirow{2}{*}{InceptionNeXt} & UPGD & 87.81 & 96.81 & 87.72 & 96.57 & 87.06 & 2.37\\
~ & +ABI & 87.06 & 96.86 & 86.93 & 96.87 & 86.50 & 92.02\\
\bottomrule
\end{tabular}
\end{table}

\begin{figure*}
\centering
\includegraphics[width = 1\linewidth]{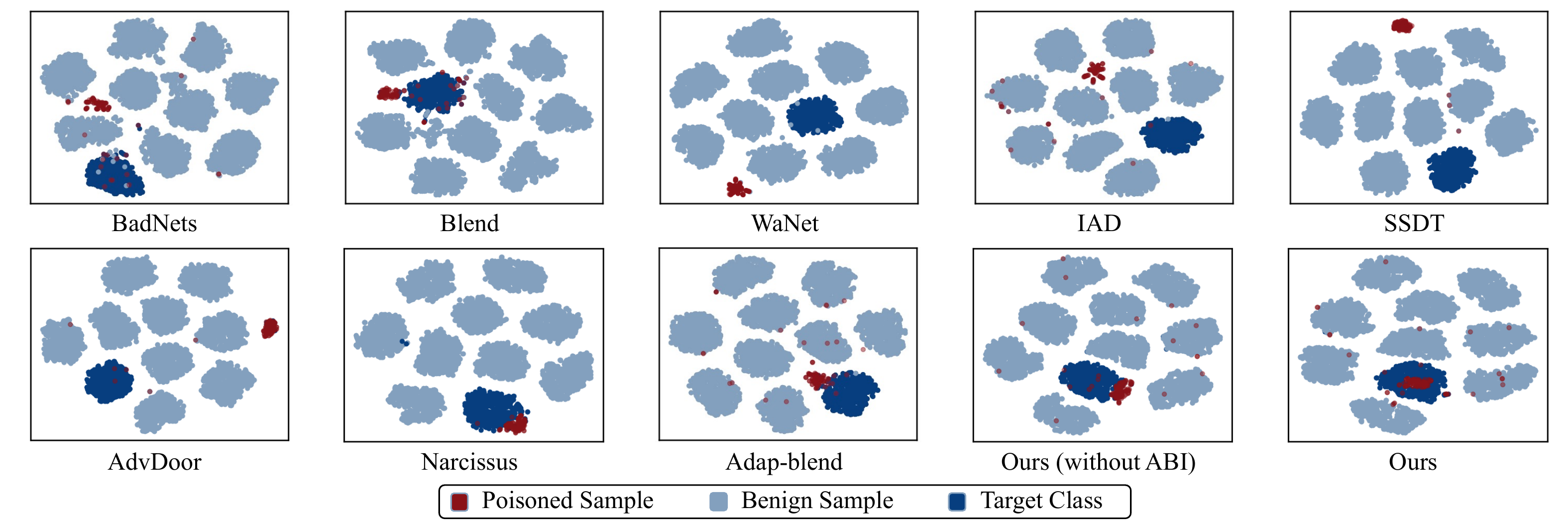}
\caption{Examples of feature visualization of \ourmethod and baseline attacks.}
\label{fig:feature_tsne}
\end{figure*}

\subsection{Ablation Study}
\label{sec:ablation-study}
The ablation study is designed for the two components of \ourmethod, the UPGD trigger and Adversarial Backdoor Injection. In addition, we also evaluate the dirty-label setting of \ourmethod and show the difference compared to using clean-label in \appcite{appendix: dirty_label}.

\noindent
\textbf{Trigger generation.}
To explore the influence of trigger patterns, we employ and evaluate three types of triggers: random noise, PGD perturbation, and UPGD perturbation, using ResNet18 on CIFAR10.
The random noise is sampled from a uniform distribution, and the PGD employs a projected gradient descent to generate sample-wise perturbations~\cite{madry2018pgd}.
The generation of UPGD is described in Algorithm~\ref{alg:upgd}.
All three triggers are limited to $8/255$ ($l_{\infty}$ norm) for imperceptibility and use the same training settings described in Table~\ref{tab:attack_config} in \appcite{appendix:back_attacks}.

Table~\ref{tab:diff_trigger} shows that random noise is ineffective as a backdoor trigger due to low ASR, even if no defense is applied.
The sample-wise PGD perturbation is more effective than random noise and shows (limited) robustness against CLP and FT-SAM.
UPGD generates the most effective backdoor trigger with an ASR higher than 80\% after CLP and FT-SAM, and we speculate that the reason is that UPGD exploits features from the target class, similar to Narcissus~\cite{zeng2023narcisus}.

Concerning exploiting features from the target class, we also explore using the natural image as the trigger, which directly contains the semantic information. 
Table~\ref {tab:semantic_trigger} shows the performance when using an automobile image from the CIFAR10 dataset as a semantic trigger.
Specifically, inspired by naturally occurring backdoor~\cite{khaddaj2023rethinking}, we design the semantic trigger by resizing the automobile image to $8\times8$ and sticking it on a part of (PR=5\%) the training images.
We consider four types of strategies, including only poisoning the same class as the trigger in clean-label, poisoning a different class (class airplane) in clean-label, only poisoning the same class as the trigger in dirty-label, and poisoning a different class (class airplane) in dirty-label.
In Table~\ref {tab:semantic_trigger}, the semantic trigger can be effective as a backdoor trigger and resist against the CLP pruning.
However, the semantic trigger is not robust against the FT-SAM fine-tuning.
The reason is that the semantic trigger cannot effectively represent the feature of a class.
Conversely, adversarial perturbation (PGD, UPGD) acquires more representative and discriminative class features, since they can capture the correlations among different regions of the decision boundary and
easily fool most of the inputs~\cite{Moosavi2017uap}.

\noindent
\textbf{Adversarial backdoor injection is critical.}
There are two components in \ourmethod: the UPGD trigger generation and Adversarial Backdoor Injection.
We conduct an ablation study with two architectures on CIFAR10 to analyze the impact of the ABI component.
As shown in Table~\ref{tab:ablation}, after removing the ABI component, CLP or FT-SAM can defend against the clean-label attack with the UPGD trigger.
Thus, the Adversarial Backdoor Injection is the key component in maintaining the effectiveness of backdoor attacks against parameter-space defenses.

\begin{table}
\centering
\caption{Evaluation with the proactive defense, CT~\cite{qi2023proactivedetection}, under different poisoning rates (PR).}
\label{tab:proactive_defense}
\begin{tabular}{cccccccccc}
\toprule
Attack & PR & ACC & ASR & Recall & FPR\\
\midrule
\multirow{5}{*}{\tabincell{c}{BadNets \\ \cite{Gu_2019_badnet}}} & 5\% & 93.18 & 99.96 & 2500/2500 & 1568/47500\\
~ & 2.5\% & 93.35 & 99.83 & 1250/1250 & 518/48750\\
~ & 1\% & 93.30 & 100 & 500/500 & 73/49500\\
~ & 0.5\% & 93.43 & 100 & 250/250 & 5/49750\\
~ & 0.3\% & 93.63 & 99.94 & 150/150 & 222/49850\\
\midrule
\multirow{5}{*}{\tabincell{c}{Adap-patch \\ \cite{qi2022revisitingadaptive}}} & 5\% & 93.28 & 100 & 1808/2500 & 116/47500\\
~ & 2.5\% & 93.68 & 100 & 1088/1250 & 20/48750\\
~ & 1\% & 93.73 & 100 & 494/500 & 570/49500\\
~ & 0.5\% & 93.31 & 100 & 160/250 & 154/49750\\
~ & 0.3\% & 93.26 & 100 & 86/150 & 3825/49850\\
\midrule
\multirow{5}{*}{\ourmethod} & 5\% & 93.84 & 99.41 & 2499/2500 & 671/47500\\
~ & 2.5\% & 93.81 & 95.83 & 115/1250 & 7220/48750\\
~ & 1\% & 94.09 & 92.48 & 208/500 & 6690/49500\\
~ & 0.5\% & 94.36 & 92.91 & 90/250 & 6738/49750\\
~ & 0.3\% & 94.22 & 90.10 & 29/150 & 6349/49850\\
\bottomrule
\end{tabular}
\end{table}

\begin{figure}
\centering
\includegraphics[width = 0.8\linewidth]{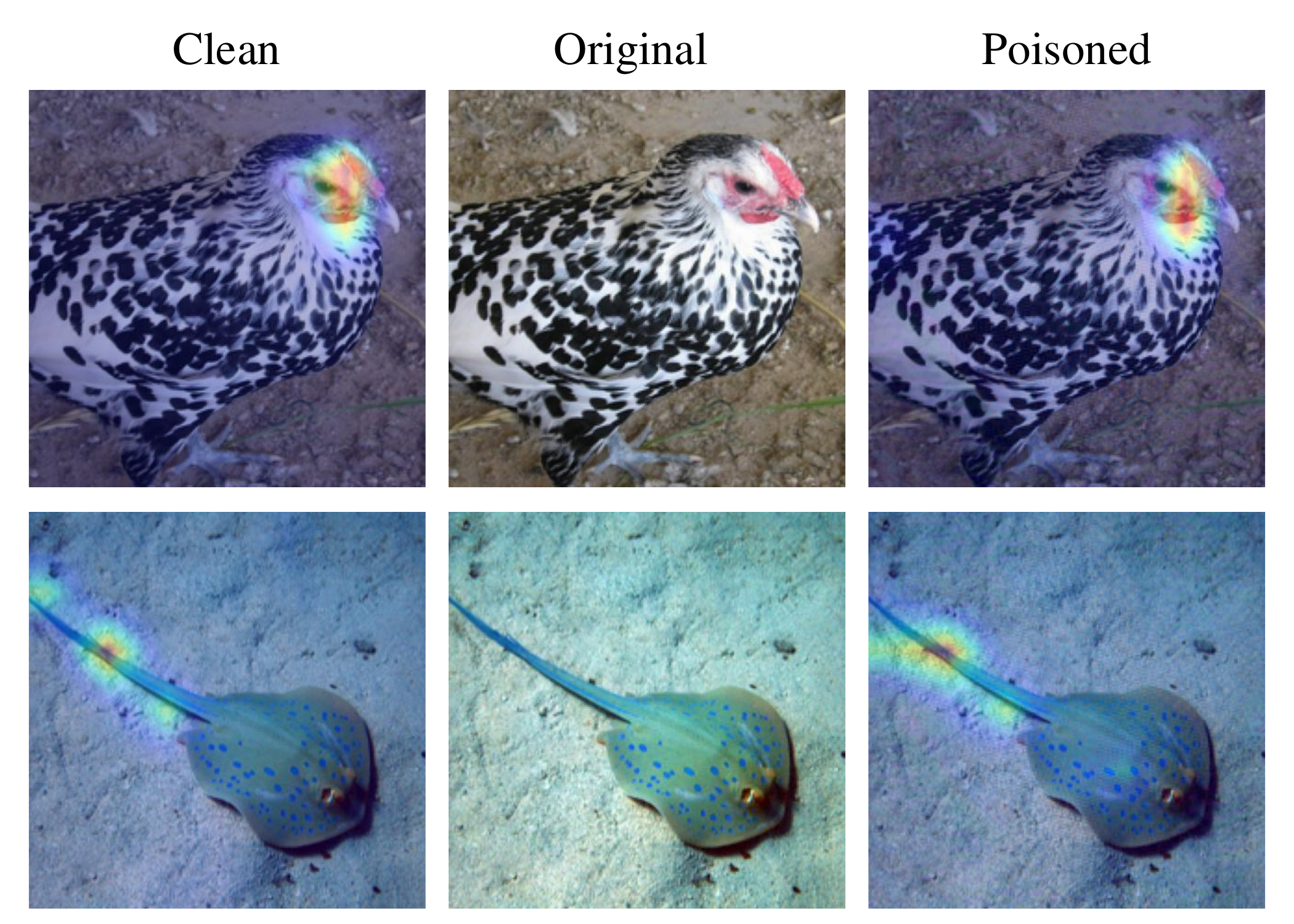}
\caption{Examples of Grad-CAM activation map with ImageNet200 images by clean and \ourmethod models. The first column is Grad-CAM maps with clean images, and the third column is Grad-CAM maps with \ourmethod-poisoned images.}
\label{fig:gradcam_images}
\end{figure}

\section{Stronger Defenders and Additional Analysis}
\label{sec:strongerdefender}

\subsection{Proactive Defense}
\label{sec:proactive}

Real-world powerful defenders could take more initiative by intervening proactively in the attack process and exploiting poisoned data. 
We evaluate \ourmethod against the SOTA proactive defense, CT~\cite{qi2023proactivedetection}, that detects poisoned samples in the training data.
Specifically, CT considers data from the original poisoned training data as regular batches and introduces randomly labeled benign data as confusing batches.
Then, CT performs normal supervised training on both regular and confusing batches to produce an inference model, aiming to corrupt benign semantic features and correlations with correct labels in the inference model by confusing batches.
The backdoor effect remains in the inference model because there is no trigger information in the confusing batches, and correctly predicted samples by the inference model are recorded as poisoned.

Table~\ref{tab:proactive_defense} presents the detection results on two baseline attacks (BadNets~\cite{Gu_2019_badnet} and Adap-patch~\cite{qi2022revisitingadaptive}) and \ourmethod.
CT is effective against the two baseline attacks, where most poisoned samples in the training set are detected with a relatively low false positive rate.
However, CT is not capable of detecting \ourmethod when the poisoning rate is lower than 5\% due to a high false positive rate and low recall.
To understand why CT is not effective against \ourmethod, we recall that the main idea of CT is to corrupt benign semantic features and their correct label, but not corrupt backdoor semantic features.
However, \ourmethod utilizes the benign semantic features of the target class to generate UPGD perturbation as the trigger.
CT's mechanism also corrupts the backdoor features of \ourmethod.
Therefore, CT cannot effectively detect \ourmethod poisoned samples.

\subsection{Visualization}
\label{sec:vis}

\noindent
\textbf{Grad-CAM cannot spot the trigger area of \ourmethod.}
Grad-CAM~\cite{Selvaraju2017gradcam} was originally designed to visualize the network's preference when taking an input image.
In backdoor defense research, Grad-CAM is leveraged to highlight the important areas in order to detect the potential backdoor trigger area~\cite{xu2020defendingbygradcam}.
Figure~\ref{fig:gradcam_images} shows the activated area of a clean model and \ourmethod backdoored model using Grad-CAM.
The activated area of \ourmethod backdoored model is indistinguishable from the clean model, so the Grad-CAM-based defense~\cite{xu2020defendingbygradcam} is also ineffective against \ourmethod.

\noindent
\textbf{t-SNE visualization of feature space.}
Figure~\ref{fig:feature_tsne} shows the latent feature (feature space of the last convolutional layers) from \ourmethod backdoor models with and without adversarial backdoor injection in 2-D space and other baseline attacks by t-SNE~\cite{van2008tsne}.
The poisoning rate for all is 0.5\%.
WaNet cannot achieve satisfactory ASR at this very low poisoning rate, so we use the default setting according to their open-source implementation.
Specifically, we perform dimensionality reduction for the latent features by t-SNE.
The model architecture is ResNet18 and trained on CIFAR10. 
Each class of samples forms a tight cluster, and \ourmethod poisoned samples are better mixed with the target class samples when the model is trained with adversarial backdoor injection.

\section{Conclusions \& Future Work}

This paper studies whether backdoor attacks can resist diverse practical defenses and provides an affirmative answer: current common stealthy backdoor attacks are vulnerable to parameter-space defenses.
We further explore how to increase the stealthiness of backdoor attacks against parameter-space defenses. 
We propose a novel supply-chain backdoor attack, \ourmethod, that considers comprehensive stealthiness, including input, feature, and parameter-space stealthiness. 
\ourmethod achieves state-of-the-art performance by leveraging adversarial examples and adaptively limiting the backdoored model's parameter changes during the backdoor injection to improve the stealthiness. 
We also show that \ourmethod's Adversarial Backdoor Injection can consistently improve other backdoor attacks against parameter space defenses.
We suggest that future backdoor attacks should be evaluated against parameter-space defense.
We also recommend that backdoor research explore Adversarial Backdoor Injection to enhance parameter-space stealthiness.

\let\clearpage\oldclearpage

\bibliographystyle{ACM-Reference-Format}
\bibliography{ref}

@inproceedings{Gu_2019_badnet,
  author={Gu, Tianyu and Liu, Kang and Dolan-Gavitt, Brendan and Garg, Siddharth},
  journal={IEEE Access}, 
  title={BadNets: Evaluating Backdooring Attacks on Deep Neural Networks}, 
  year={2019},
}

@inproceedings{liu_2018_Trojannn,
  author    = {Yingqi Liu and
               Shiqing Ma and
               Yousra Aafer and
               Wen-Chuan Lee and
               Juan Zhai and
               Weihang Wang and
               Xiangyu Zhang},
  title     = {Trojaning Attack on Neural Networks},
  booktitle = {Network and Distributed System Security (NDSS) Symposium},
  year      = {2018},
}

@inproceedings{chen_2017_blend,
      title={Targeted Backdoor Attacks on Deep Learning Systems Using Data Poisoning}, 
      author={Xinyun Chen and Chang Liu and Bo Li and Kimberly Lu and Dawn Song},
      year={2017},
      journal={arXiv preprint arXiv:1712.05526}
}

@inproceedings{nguyen2021wanet,
title={WaNet - Imperceptible Warping-based Backdoor Attack},
author={Tuan Anh Nguyen and Anh Tuan Tran},
booktitle={International Conference on Learning Representations (ICLR)},
year={2021},
}

@inproceedings{nguyen2020inputaware,
  title={Input-Aware Dynamic Backdoor Attack},
  author={Nguyen, Anh and Tran, Anh},
  booktitle={Advances in Neural Information Processing Systems (NeurIPS)},
  year={2020}
}

@inproceedings{Wang_2022_bpp,
    author    = {Wang, Zhenting and Zhai, Juan and Ma, Shiqing},
    title     = {BppAttack: Stealthy and Efficient Trojan Attacks Against Deep Neural Networks via Image Quantization and Contrastive Adversarial Learning},
    booktitle = {IEEE / CVF Computer Vision and Pattern Recognition Conference (CVPR)},
    year      = {2022}
}

@inproceedings {Bagdasaryan_2021_blind,
author = {Eugene Bagdasaryan and Vitaly Shmatikov},
title = {Blind Backdoors in Deep Learning Models},
booktitle = {USENIX Security Symposium},
year = {2021},
}

@inproceedings{chen2018detecting,
      title={Detecting Backdoor Attacks on Deep Neural Networks by Activation Clustering}, 
      author={Bryant Chen and Wilka Carvalho and Nathalie Baracaldo and Heiko Ludwig and Benjamin Edwards and Taesung Lee and Ian Molloy and Biplav Srivastava},
      year={2018},
      journal={SafeAI Workshop @ AAAI}
}

@inproceedings{Wu_2021_anp,
 author = {Wu, Dongxian and Wang, Yisen},
 booktitle = {Advances in Neural Information Processing Systems (NeurIPS)},
 title = {Adversarial Neuron Pruning Purifies Backdoored Deep Models},
 year = {2021}
}

@inproceedings{wang_2019_nc,
  author={Wang, Bolun and Yao, Yuanshun and Shan, Shawn and Li, Huiying and Viswanath, Bimal and Zheng, Haitao and Zhao, Ben Y.},
  booktitle={IEEE Symposium on
Security and Privacy (S\&P)}, 
  title={Neural Cleanse: Identifying and Mitigating Backdoor Attacks in Neural Networks}, 
  year={2019},
}

@inproceedings{liu_2019_abs,
author = {Liu, Yingqi and Lee, Wen-Chuan and Tao, Guanhong and Ma, Shiqing and Aafer, Yousra and Zhang, Xiangyu},
title = {ABS: Scanning Neural Networks for Back-doors by Artificial Brain Stimulation},
year = {2019},
booktitle = {ACM Conference on Computer and Communications Security (CCS)},

}

@inproceedings{wang2022featurere,
  title={Rethinking the Reverse-engineering of Trojan Triggers},
  author={Wang, Zhenting and Mei, Kai and Ding, Hailun and Zhai, Juan and Ma, Shiqing},
  booktitle={Advances in Neural Information Processing Systems (NeurIPS)},
  year={2022}
}

@inproceedings{
wang2023unicorn,
title={{UNICORN}: A Unified Backdoor Trigger Inversion Framework},
author={Zhenting Wang and Kai Mei and Juan Zhai and Shiqing Ma},
booktitle={International Conference on Learning Representations (ICLR)},
year={2023},
}

@inproceedings{xu2024btidbf,
  title={Towards Reliable and Efficient Backdoor Trigger Inversion via Decoupling Benign Features},
  author={Xu, Xiong and Huang, Kunzhe and Li, Yiming and Qin, Zhan and Ren, Kui},
  booktitle={International Conference on Learning Representations (ICLR)},
  year={2024}
}

@inproceedings{
madry2018pgd,
title={Towards Deep Learning Models Resistant to Adversarial Attacks},
author={Aleksander Madry and Aleksandar Makelov and Ludwig Schmidt and Dimitris Tsipras and Adrian Vladu},
booktitle={International Conference on Learning Representations (ICLR)},
year={2018},
}

@inproceedings{guo_2020_tabor,
  author={Guo, Wenbo and Wang, Lun and Xu, Yan and Xing, Xinyu and Du, Min and Song, Dawn},
  booktitle={IEEE International Conference on Data Mining (ICDM)}, 
  title={Towards Inspecting and Eliminating Trojan Backdoors in Deep Neural Networks}, 
  year={2020},
}

@inproceedings{qi2022revisitingadaptive,
  title={Revisiting the assumption of latent separability for backdoor defenses},
  author={Qi, Xiangyu and Xie, Tinghao and Li, Yiming and Mahloujifar, Saeed and Mittal, Prateek},
  booktitle={International Conference on Learning Representations (ICLR)},
  year={2023}
}

@inproceedings{mo2024ssdt,
  title={Robust Backdoor Detection for Deep Learning via Topological Evolution Dynamics},
  author={Mo, Xiaoxing and Zhang, Yechao and Zhang, Leo Yu and Luo, Wei and Sun, Nan and Hu, Shengshan and Gao, Shang and Xiang, Yang},
  booktitle={IEEE Symposium on
Security and Privacy (S\&P)},
  year={2024}
}

@inproceedings{krizhevsky2009cifar10,
  title={Learning multiple layers of features from tiny images},
  author={Krizhevsky, Alex and Hinton, Geoffrey and others},
  year={2009},
  journal={Technical Report, University of Toronto}
}

@inproceedings{jia2009imagenet,
  author={Deng, Jia and Dong, Wei and Socher, Richard and Li, Li-Jia and Kai Li and Li Fei-Fei},
  booktitle={IEEE / CVF Computer Vision and Pattern Recognition Conference (CVPR)}, 
  title={ImageNet: A large-scale hierarchical image database}, 
  year={2009}
}

@inproceedings{He_2016_resnet,
author = {He, Kaiming and Zhang, Xiangyu and Ren, Shaoqing and Sun, Jian},
title = {Deep Residual Learning for Image Recognition},
booktitle = {IEEE / CVF Computer Vision and Pattern Recognition Conference (CVPR)},
year = {2016}
}

@inproceedings{Karen_2015_vgg,
  author       = {Karen Simonyan and
                  Andrew Zisserman},
  title        = {Very Deep Convolutional Networks for Large-Scale Image Recognition},
  booktitle    = {International Conference on Learning Representations (ICLR)},
  year         = {2015},
}

@inproceedings{huang2017densely,
  title={Densely connected convolutional networks},
  author={Huang, Gao and Liu, Zhuang and Van Der Maaten, Laurens and Weinberger, Kilian Q},
  booktitle={IEEE / CVF Computer Vision and Pattern Recognition Conference (CVPR)},
  year={2017}
}

@inproceedings{pang_2022_trojanzoo,
      title={TrojanZoo: Towards Unified, Holistic, and Practical Evaluation of Neural Backdoors}, 
      author={Ren Pang and Zheng Zhang and Xiangshan Gao and Zhaohan Xi and Shouling Ji and Peng Cheng and Ting Wang},
      year={2022},
      booktitle={IEEE Symposium on Security and Privacy (Euro S\&P)},
}

@inproceedings{Stallkamp2012gtsrb,
title = "Man vs. computer: Benchmarking machine learning algorithms for traffic sign recognition",
journal = "Neural Networks",
year = "2012",
author = "J. Stallkamp and M. Schlipsing and J. Salmen and C. Igel"
}

@inproceedings{hong2022handcrafted,
  title={Handcrafted backdoors in deep neural networks},
  author={Hong, Sanghyun and Carlini, Nicholas and Kurakin, Alexey},
  journal={Advances in Neural Information Processing Systems (NeurIPS)},
  year={2022}
}

@inproceedings{
guo2023scaleup,
title={{SCALE}-{UP}: An Efficient Black-box Input-level Backdoor Detection via Analyzing Scaled Prediction Consistency},
author={Junfeng Guo and Yiming Li and Xun Chen and Hanqing Guo and Lichao Sun and Cong Liu},
booktitle={International Conference on Learning Representations (ICLR)},
year={2023}
}

@inproceedings{zeng2023narcisus,
author = {Zeng, Yi and Pan, Minzhou and Just, Hoang Anh and Lyu, Lingjuan and Qiu, Meikang and Jia, Ruoxi},
title = {Narcissus: A Practical Clean-Label Backdoor Attack with Limited Information},
year = {2023},
booktitle = {ACM Conference on Computer and Communications Security (CCS)}
}

@inproceedings{turner2019lc,
  title={Label-consistent backdoor attacks},
  author={Turner, Alexander and Tsipras, Dimitris and Madry, Aleksander},
  journal={arXiv preprint arXiv:1912.02771},
  year={2019}
}

@inproceedings{liu2018finepruning,
author="Liu, Kang
and Dolan-Gavitt, Brendan
and Garg, Siddharth",
title="Fine-Pruning: Defending Against Backdooring Attacks on Deep Neural Networks",
booktitle="Research in Attacks, Intrusions, and Defenses",
year="2018",
}

@inproceedings{zheng2022clp,
author="Zheng, Runkai
and Tang, Rongjun
and Li, Jianze
and Liu, Li",
title="Data-Free Backdoor Removal Based on Channel Lipschitzness",
booktitle="European Conference on Computer Vision (ECCV)",
year="2022",
}

@inproceedings{
li2023rnp,
title={Reconstructive Neuron Pruning for Backdoor Defense},
author={Yige Li and Xixiang Lyu and Xingjun Ma and Nodens Koren and Lingjuan Lyu and Bo Li and Yu-Gang Jiang},
booktitle={International Conference on Machine Learning (ICML)},
year={2023},
}

@inproceedings{Zhu2023ftsam,
    author    = {Zhu, Mingli and Wei, Shaokui and Shen, Li and Fan, Yanbo and Wu, Baoyuan},
    title     = {Enhancing Fine-Tuning Based Backdoor Defense with Sharpness-Aware Minimization},
    booktitle = {International Conference on Computer Vision (ICCV)},
    year      = {2023},
}

@inproceedings{min2023fst,
 author = {Min, Rui and Qin, Zeyu and Shen, Li and Cheng, Minhao},
 booktitle = {Advances in Neural Information Processing Systems (NeurIPS)},
 title = {Towards Stable Backdoor Purification through Feature Shift Tuning},
 year = {2023}
}

@inproceedings{
zeng2022ibau,
title={Adversarial Unlearning of Backdoors via Implicit Hypergradient},
author={Yi Zeng and Si Chen and Won Park and Zhuoqing Mao and Ming Jin and Ruoxi Jia},
booktitle={International Conference on Learning Representations (ICLR)},
year={2022}
}

@inproceedings{Moosavi2017uap,
author = {Moosavi-Dezfooli, Seyed-Mohsen and Fawzi, Alhussein and Fawzi, Omar and Frossard, Pascal},
title = {Universal Adversarial Perturbations},
booktitle = {IEEE / CVF Computer Vision and Pattern Recognition Conference (CVPR)},
year = {2017}
}

@inproceedings{lin2024unveiling,
      title={Unveiling and Mitigating Backdoor Vulnerabilities based on Unlearning Weight Changes and Backdoor Activeness}, 
      author={Weilin Lin and Li Liu and Shaokui Wei and Jianze Li and Hui Xiong},
      year={2024},
      booktitle={Advances in Neural Information Processing Systems (NeurIPS)}
}

@inproceedings{zhang2021advdoor,
author = {Zhang, Quan and Ding, Yifeng and Tian, Yongqiang and Guo, Jianmin and Yuan, Min and Jiang, Yu},
title = {AdvDoor: adversarial backdoor attack of deep learning system},
year = {2021},
booktitle = {ACM SIGSOFT International Symposium on Software Testing and Analysis (ISSTA)}
}

@inproceedings{abad2024gradientshaping,
  title={Gradient Shaping: Enhancing Backdoor Attack Against Reverse Engineering},
  author={Zhu, Rui and Tang, Di and Tang, Siyuan and Tao, Guanhong and Ma, Shiqing and Wang, Xiaofeng and Tang, Haixu},
  booktitle={Network and Distributed System Security (NDSS) Symposium},
  year={2024}
}

@inproceedings{shokri2020bypassing,
  title={Bypassing backdoor detection algorithms in deep learning},
  author={Shokri, Reza and others},
  booktitle={IEEE Symposium on Security and Privacy (Euro S\&P)},
  year={2020}
}

@inproceedings{
xu2024ban,
title={{BAN}: Detecting Backdoors Activated by Neuron Noise},
author={Xiaoyun Xu and Zhuoran Liu and Stefanos Koffas and Shujian Yu and Stjepan Picek},
booktitle={Advances in Neural Information Processing Systems (NeurIPS)},
year={2024}
}

@InProceedings{tan2019efficientnet,
  title = 	 {{E}fficient{N}et: Rethinking Model Scaling for Convolutional Neural Networks},
  author =       {Tan, Mingxing and Le, Quoc},
  booktitle = 	 {International Conference on Machine Learning (ICML)},
  year = 	 {2019},
}

@inproceedings{Yu2024inceptionnext,
    author    = {Yu, Weihao and Zhou, Pan and Yan, Shuicheng and Wang, Xinchao},
    title     = {InceptionNeXt: When Inception Meets ConvNeXt},
    booktitle = {IEEE / CVF Computer Vision and Pattern Recognition Conference (CVPR)},
    year      = {2024}
}

@inproceedings{liu2017faultinjection,
  author={Liu, Yannan and Wei, Lingxiao and Luo, Bo and Xu, Qiang},
  booktitle={IEEE International Conference on Computer-Aided Design (ICCAD)}, 
  title={Fault injection attack on deep neural network}, 
  year={2017}
}

@inproceedings{Qi2022deploymentbackdoor,
    author    = {Qi, Xiangyu and Xie, Tinghao and Pan, Ruizhe and Zhu, Jifeng and Yang, Yong and Bu, Kai},
    title     = {Towards Practical Deployment-Stage Backdoor Attack on Deep Neural Networks},
    booktitle = {IEEE / CVF Computer Vision and Pattern Recognition Conference (CVPR)},
    year      = {2022}
}

@inproceedings{xia2023mmdregularization,
  author={Xia, Pengfei and Niu, Hongjing and Li, Ziqiang and Li, Bin},
  journal={IEEE Transactions on Dependable and Secure Computing (TDSC)}, 
  title={Enhancing Backdoor Attacks With Multi-Level MMD Regularization}, 
  year={2023}
}

@inproceedings{doan2021wb,
 author = {Doan, Khoa and Lao, Yingjie and Li, Ping},
 booktitle = {Advances in Neural Information Processing Systems (NeurIPS)},
 title = {Backdoor Attack with Imperceptible Input and Latent Modification},
 year = {2021}
}

@inproceedings{ren2021Simtrojan,
  author={Ren, Yankun and Li, Longfei and Zhou, Jun},
  booktitle={IEEE International Conference on Image Processing (ICIP)}, 
  title={Simtrojan: Stealthy Backdoor Attack}, 
  year={2021},
}

@inproceedings{Cheng2021dfst, 
title={Deep Feature Space Trojan Attack of Neural Networks by Controlled Detoxification},
journal={AAAI Conference on Artificial Intelligence (AAAI)}, author={Cheng, Siyuan and Liu, Yingqi and Ma, Shiqing and Zhang, Xiangyu}, year={2021} }

@inproceedings{nan2022iba,
  author       = {Nan Zhong and
                  Zhenxing Qian and
                  Xinpeng Zhang},
  title        = {Imperceptible Backdoor Attack: From Input Space to Feature Representation},
  booktitle    = {International Joint Conferences on Artificial Intelligence (IJCAI)},
  year         = {2022},
}

@inproceedings{zhao2022defeat,
    author    = {Zhao, Zhendong and Chen, Xiaojun and Xuan, Yuexin and Dong, Ye and Wang, Dakui and Liang, Kaitai},
    title     = {DEFEAT: Deep Hidden Feature Backdoor Attacks by Imperceptible Perturbation and Latent Representation Constraints},
    booktitle = {IEEE / CVF Computer Vision and Pattern Recognition Conference (CVPR)},
    year      = {2022}
}

@inproceedings {di2021tact,
author = {Di Tang and XiaoFeng Wang and Haixu Tang and Kehuan Zhang},
title = {Demon in the Variant: Statistical Analysis of {DNNs} for Robust Backdoor Contamination Detection},
booktitle = {USENIX Security Symposium},
year = {2021}
}

@inproceedings{li2021abl,
 author = {Li, Yige and Lyu, Xixiang and Koren, Nodens and Lyu, Lingjuan and Li, Bo and Ma, Xingjun},
 booktitle = {Advances in Neural Information Processing Systems (NeurIPS)},
 title = {Anti-Backdoor Learning: Training Clean Models on Poisoned Data},
 year = {2021}
}

@inproceedings{gao2019strip,
author = {Gao, Yansong and Xu, Change and Wang, Derui and Chen, Shiping and Ranasinghe, Damith C. and Nepal, Surya},
title = {STRIP: a defence against trojan attacks on deep neural networks},
year = {2019},
booktitle = {Annual Computer Security Applications Conference (ACSAC)},
}

@inproceedings{foret2021sam,
title={Sharpness-aware Minimization for Efficiently Improving Generalization},
author={Pierre Foret and Ariel Kleiner and Hossein Mobahi and Behnam Neyshabur},
booktitle={International Conference on Learning Representations (ICLR)},
year={2021}
}

@inproceedings{
hou2024ibdpsc,
title={{IBD}-{PSC}: Input-level Backdoor Detection via Parameter-oriented Scaling Consistency},
author={Linshan Hou and Ruili Feng and Zhongyun Hua and Wei Luo and Leo Yu Zhang and Yiming Li},
booktitle={International Conference on Machine Learning (ICML)},
year={2024}
}

@inproceedings{wu2022backdoorbench,
 author = {Wu, Baoyuan and Chen, Hongrui and Zhang, Mingda and Zhu, Zihao and Wei, Shaokui and Yuan, Danni and Shen, Chao},
 booktitle = {Advances in Neural Information Processing Systems (NeurIPS)},
 title = {BackdoorBench: A Comprehensive Benchmark of Backdoor Learning},
 year = {2022}
}

@inproceedings{li2023backdoorbox,
  title={{BackdoorBox}: A Python Toolbox for Backdoor Learning},
  author={Li, Yiming and Ya, Mengxi and Bai, Yang and Jiang, Yong and Xia, Shu-Tao},
  booktitle={International Conference on Learning Representations (ICLR) Workshop},
  year={2023}
}

@inproceedings{
tsipras2018robustnessoddwithacc,
title={Robustness May Be at Odds with Accuracy},
author={Dimitris Tsipras and Shibani Santurkar and Logan Engstrom and Alexander Turner and Aleksander Madry},
booktitle={International Conference on Learning Representations (ICLR)},
year={2019}
}

@inproceedings{Selvaraju2017gradcam,
author = {Selvaraju, Ramprasaath R. and Cogswell, Michael and Das, Abhishek and Vedantam, Ramakrishna and Parikh, Devi and Batra, Dhruv},
title = {Grad-CAM: Visual Explanations From Deep Networks via Gradient-Based Localization},
booktitle = {International Conference on Computer Vision (ICCV)},
year = {2017}
}

@inproceedings{van2008tsne,
  title={Visualizing data using t-SNE},
  author={Van der Maaten, Laurens and Hinton, Geoffrey},
  journal={Journal of Machine Learning Research (JMLR)},
  year={2008}
}

@inproceedings{
cao2024dfba,
title={Data Free Backdoor Attacks},
author={Bochuan Cao and Jinyuan Jia and Chuxuan Hu and Wenbo Guo and Zhen Xiang and Jinghui Chen and Bo Li and Dawn Song},
booktitle={Advances in Neural Information Processing Systems (NeurIPS)},
year={2024}
}

@inproceedings{
wei2024pdb,
title={Mitigating Backdoor Attack by Injecting Proactive Defensive Backdoor},
author={Shaokui Wei and Hongyuan Zha and Baoyuan Wu},
booktitle={Advances in Neural Information Processing Systems (NeurIPS)},
year={2024}
}

@inproceedings {qi2023proactivedetection,
author = {Xiangyu Qi and Tinghao Xie and Jiachen T. Wang and Tong Wu and Saeed Mahloujifar and Prateek Mittal},
title = {Towards A Proactive {ML} Approach for Detecting Backdoor Poison Samples},
booktitle = {USENIX Security Symposium},
year = {2023}
}

@inproceedings{liu2020compositeattack,
author = {Lin, Junyu and Xu, Lei and Liu, Yingqi and Zhang, Xiangyu},
title = {Composite Backdoor Attack for Deep Neural Network by Mixing Existing Benign Features},
year = {2020},
booktitle = {ACM Conference on Computer and Communications Security (CCS)},
}

@inproceedings{rakin2022tbfa,
  author={Rakin, Adnan Siraj and He, Zhezhi and Li, Jingtao and Yao, Fan and Chakrabarti, Chaitali and Fan, Deliang},
  journal={IEEE Transactions on Pattern Analysis and Machine Intelligence (TPAMI)}, 
  title={T-BFA: Targeted Bit-Flip Adversarial Weight Attack}, 
  year={2022}
}

@inproceedings{Rakin_2020_TBT,
author = {Rakin, Adnan Siraj and He, Zhezhi and Fan, Deliang},
title = {TBT: Targeted Neural Network Attack With Bit Trojan},
booktitle = {IEEE / CVF Computer Vision and Pattern Recognition Conference (CVPR)},
year = {2020}
}

@inproceedings{chen2021proflip,
    author    = {Chen, Huili and Fu, Cheng and Zhao, Jishen and Koushanfar, Farinaz},
    title     = {ProFlip: Targeted Trojan Attack With Progressive Bit Flips},
    booktitle = {International Conference on Computer Vision (ICCV)},
    year      = {2021}
}

@inproceedings {lv2023datafree,
author = {Peizhuo Lv and Chang Yue and Ruigang Liang and Yunfei Yang and Shengzhi Zhang and Hualong Ma and Kai Chen},
title = {A Data-free Backdoor Injection Approach in Neural Networks},
booktitle = {USENIX Security Symposium},
year = {2023}
}

@inproceedings{xu2020defendingbygradcam,
  title={Defending against backdoor attack on deep neural networks},
  author={Hao Cheng and Kaidi Xu and Sijia Liu and Pin-Yu Chen and Pu Zhao and Xue Lin},
  journal={AdvML Workshop @ KDD 2019},
  year={2020}
}

@InProceedings{Doan2021lira,
    author    = {Doan, Khoa and Lao, Yingjie and Zhao, Weijie and Li, Ping},
    title     = {LIRA: Learnable, Imperceptible and Robust Backdoor Attacks},
    booktitle = {International Conference on Computer Vision (ICCV)},
    year      = {2021}
}

@InProceedings{Liu2021swin,
    author    = {Liu, Ze and Lin, Yutong and Cao, Yue and Hu, Han and Wei, Yixuan and Zhang, Zheng and Lin, Stephen and Guo, Baining},
    title     = {Swin Transformer: Hierarchical Vision Transformer Using Shifted Windows},
    booktitle = {International Conference on Computer Vision (ICCV)},
    year      = {2021}
}

@inproceedings{
dosovitskiy2021vit,
title={An Image is Worth 16x16 Words: Transformers for Image Recognition at Scale},
author={Alexey Dosovitskiy and Lucas Beyer and Alexander Kolesnikov and Dirk Weissenborn and Xiaohua Zhai and Thomas Unterthiner and Mostafa Dehghani and Matthias Minderer and Georg Heigold and Sylvain Gelly and Jakob Uszkoreit and Neil Houlsby},
booktitle={International Conference on Learning Representations (ICLR)},
year={2021}
}

@InProceedings{khaddaj2023rethinking,
  title = 	 {Rethinking Backdoor Attacks},
  author =       {Khaddaj, Alaa and Leclerc, Guillaume and Makelov, Aleksandar and Georgiev, Kristian and Salman, Hadi and Ilyas, Andrew and Madry, Aleksander},
  booktitle = 	 {International Conference on Machine Learning (ICML)},
  year = 	 {2023}
}

@article{xu2025towards,
  title={Towards Backdoor Stealthiness in Model Parameter Space},
  author={Xu, Xiaoyun and Liu, Zhuoran and Koffas, Stefanos and Picek, Stjepan},
  journal={arXiv preprint arXiv:2501.05928},
  year={2025}
}

\appendix
\section*{Appendix}
Full version with appendix:~\cite{xu2025towards}.

\section{Additional Details about Experimental Settings}

Complete appendix and more experimental results can be found in the arXiv version of the paper.

\subsection{Datasets}
\label{appendix:datasets}

\noindent
\textbf{CIFAR10.}
The CIFAR10~\cite{krizhevsky2009cifar10} contains 50,000 training images and 10,000 testing images with the size of $3\times32\times32$ in 10 classes.

\noindent
\textbf{GTSRB.}
The GTSRB~\cite{Stallkamp2012gtsrb} contains 39,209 training images and 12,630 testing images in 43 classes. In our experiments, the images are resized to $3\times32\times32$.

\noindent
\textbf{ImageNet200.}
ImageNet~\cite{jia2009imagenet} contains over 1.2 million high-resolution images in 1,000 classes.
In our experiments, we randomly select 200 classes from the ImageNet dataset as our ImageNet200 dataset.
Each class has 1,300 training images and 50 testing images.
The ImageNet images are resized to $3\times224\times224$.

\subsection{Backdoor Attacks}
\label{appendix:back_attacks}

Our attack is compared with 12 well-known and representative attacks: BadNets~\cite{Gu_2019_badnet}, Blend~\cite{chen_2017_blend}, WaNet~\cite{nguyen2021wanet}, IAD~\cite{nguyen2020inputaware}, AdvDoor~\cite{zhang2021advdoor}, BppAttack~\cite{Wang_2022_bpp}, LC~\cite{turner2019lc}, Narcissus~\cite{zeng2023narcisus}, Adap-Blend~\cite{qi2022revisitingadaptive}, SSDT~\cite{mo2024ssdt}, DFST~\cite{Cheng2021dfst}, and DFBA~\cite{cao2024dfba}.

\begin{table}[htb]
    \centering
    \footnotesize
        \caption{The backdoor training settings.}
    \begin{tabular}{l|l}
        \toprule
         Config & Value \\
         \midrule
         \multirow{2}{*}{Optimizer} & SGD, \\ ~ & AdamW (InceptionNeXt)\\
         Weight decay & $5\times 10^{-4}$ \\
         learning rate & 0.01 \\
         epoch & 200 (GTSRB, CIFAR10), 100 (ImageNet200) \\
         \multirow{2}{*}{learning rate schedule} & MultiStepLR (100, 150) for CIFAR10 and GTSRB, \\ ~ & CosineAnnealingLR for ImageNet200\\
         poison rate & 0.05 \\
         $u$ in~\Eqref{eq:advesarial_injection} & 3.0 \\
         BadNets trigger & $3\times3$\\
         Blend trigger &  random Gaussian noise and blend ratio 0.2\\
         Adap-blend trigger &  ``hellokitty\_32.png'' and blend ratio of 0.2 \\
         Narcissus trigger size & $\epsilon=16$ for both inference and training\\ 
         \bottomrule
    \end{tabular}
    \label{tab:attack_config}
\end{table}

Like Narcissus, our attack uses the class bird (CIFAR10) as the target class.
For ImageNet200, we use the ``stingray'' as the target class.
The \ourmethod poisoning rate (ImageNet200) used for results in Table~\ref{tab:datasets_allattacks} is 0.25\%.
For GTSRB, we use the speed limit (50) as the target class.
The \ourmethod poisoning rate (GTSRB) used for results in Table~\ref{tab:datasets_allattacks} is 1.74\%.
AdvDoor uses the same trigger and target class as ours.
More details are provided in Table~\ref{tab:attack_config}.
For other attacks and hyperparameters not mentioned, we use the default setting from the original papers or open-source implementations.

\subsection{Backdoor Defenses}
\label{appendix:defenses}

We evaluate our attack and baseline attacks against 17 defenses, including \textbf{4 pruning-based methods} (FP~\cite{liu2018finepruning}, ANP~\cite{Wu_2021_anp}, CLP~\cite{zheng2022clp}, and RNP~\cite{li2023rnp}),
\textbf{5 fine-tuning-based methods} (vanilla FT, FT-SAM~\cite{Zhu2023ftsam}, I-BAU~\cite{zeng2022ibau}, FST~\cite{min2023fst}, and BTI-DBF(U)~\cite{xu2024btidbf}), \textbf{5 backdoor model detections} (NC~\cite{wang_2019_nc}, Tabor~\cite{guo_2020_tabor}, FeatureRE~\cite{wang2022featurere}, Unicorn~\cite{wang2023unicorn}, and BTI-DBF~\cite{xu2024btidbf}),  \textbf{2 backdoor input detections} (Scale-up~\cite{guo2023scaleup} and IBD-PSC~\cite{hou2024ibdpsc}), and \textbf{a proactive detection} CT~\cite{qi2023proactivedetection}.

\textbf{ANP\footnote{\url{https://github.com/csdongxian/ANP_backdoor/tree/main}}, CLP\footnote{\url{https://github.com/rkteddy/channel-Lipschitzness-based-pruning}}, RNP\footnote{\url{https://github.com/bboylyg/RNP}}, FST\footnote{\url{https://github.com/AISafety-HKUST/Backdoor_Safety_Tuning}}, BTI-DBF\footnote{\url{https://github.com/xuxiong0214/BTIDBF/tree/master}\label{bitdbffootnote}}, BTI-DBF(U)\footref{bitdbffootnote}, FeatureRE\footnote{\url{https://github.com/RU-System-Software-and-Security/FeatureRE/tree/main}}, Unicorn\footnote{\url{https://github.com/RU-System-Software-and-Security/UNICORN}}}. We use the implementation and default hyperparameters from their open-source code.

\textbf{FP\footnote{\url{https://github.com/SCLBD/BackdoorBench/blob/main/defense/fp.py}}, vanilla FT\footnote{\url{https://github.com/SCLBD/BackdoorBench/blob/main/defense/ft.py}}, FT-SAM\footnote{\url{https://github.com/SCLBD/BackdoorBench/blob/main/defense/ft-sam.py}}, I-BAU\footnote{\url{https://github.com/SCLBD/BackdoorBench/blob/main/defense/i-bau.py}}}. We use the implementation and default hyperparameters from BackdoorBench~\cite{wu2022backdoorbench}.
For FT-SAM on ImageNet200, the default setting will decrease benign accuracy to 0.465, so we reduce its training schedule to 25 epochs. 
Please note that the experiments on CIFAR10 with FT-SAM usually converge within 20 epochs in our experiments.
Thus, decreasing the training schedule is not harmful to the defense performance. 

\textbf{NC\footnote{\url{https://github.com/ain-soph/trojanzoo/blob/main/trojanvision/defenses/backdoor/model_inspection/neural_cleanse.py}} and Tabor\footnote{\url{https://github.com/ain-soph/trojanzoo/blob/main/trojanvision/defenses/backdoor/model_inspection/tabor.py}}.} We use the implementation from TrojanZoo~\cite{pang_2022_trojanzoo}. 1\% training set and 100 epochs are used for trigger inversion.

\textbf{Scale-up\footnote{\url{https://github.com/THUYimingLi/BackdoorBox/blob/main/core/defenses/SCALE_UP.py}}, IBD-PSC\footnote{\url{https://github.com/THUYimingLi/BackdoorBox/blob/main/core/defenses/IBD_PSC.py}}}. We use the implementation and default hyperparameters from BackdoorBox~\cite{li2023backdoorbox}.

\textbf{CT\footnote{\url{https://github.com/Unispac/Fight-Poison-With-Poison}}}. We use the open-source code implementation. We reduced the number of distillation iterations to 200 for efficiency reasons.

\subsection{Hyperparameters for Training Surrogate Models}
\label{appendix:training_surrogate}

Table~\ref{tab:surrogate_config} provides the hyperparameters for training surrogate models to generate UPGD.

\begin{table}[htb]
    \centering
    \footnotesize
        \caption{The settings for training surrogate models.}

    \begin{tabular}{l|l}
        \toprule
         Config & Value \\
         \midrule
         Optimizer & SGD, AdamW (InceptionNeXt)\\
         Weight decay & $5\times 10^{-4}$ \\
         learning rate & 0.01 (CIFAR10, GTSRB), 0.001 (ImageNet200) \\
         epoch & 200 (GTSRB, CIFAR10), 100 (ImageNet200) \\
         \multirow{2}{*}{learning rate schedule} & MultiStepLR (100, 150) for CIFAR10 and GTSRB, \\ ~ & CosineAnnealingLR for ImageNet200\\
         \bottomrule
    \end{tabular}
    \label{tab:surrogate_config}
\end{table}

\section{Attack Summary}
\label{appendix:threat_model}

In Table~\ref{tab:attack_summary}, we summarize the attacks evaluated in this work and compare them with \ourmethod. \ourmethod is the only one that achieves stealthiness in input, feature, and parameter spaces.

\begin{table*}[htb]
\centering
\footnotesize
\setlength\tabcolsep{3.5pt}
\begin{threeparttable}
\caption{A summary of attacks evaluated in this work. 
}
\label{tab:attack_summary}
\begin{tabular}{cccccccccccc}
\toprule
\multirow{2}{*}{Attack} & \multicolumn{3}{c}{Threat Model} & \multicolumn{3}{c}{Trigger Type} & \multicolumn{2}{c}{Trigger Strategy} & \multicolumn{3}{c}{Stealthy Level}\\
\cmidrule(lr){2-4}
\cmidrule(lr){5-7}
\cmidrule(lr){8-9}
\cmidrule(lr){10-12}
~ & Poisoning data & Clean Label & Access Training/Model & Patch & Blend & Dynamic & All-to-all & Source Class Specific & Input & Feature  & Parameter \\
\midrule
BadNets~\cite{Gu_2019_badnet} & \pie{360} & \pie{90} & \pie{90} & \pie{360} & \pie{90} & \pie{90} & \pie{360} & \pie{90} & \pie{90} & \pie{90} & \pie{90}\\
Blend~\cite{chen_2017_blend} & \pie{360} & \pie{90} & \pie{90} & \pie{90} & \pie{360} & \pie{90} & \pie{360} & \pie{90} & \pie{90} & \pie{90} & \pie{90}\\
WaNet~\cite{nguyen2021wanet} & \pie{360} & \pie{90} & \pie{90} & \pie{90} & \pie{90} & \pie{360} & \pie{360} & \pie{90} & \pie{360} & \pie{90} & \pie{90}\\
IAD~\cite{nguyen2020inputaware} & \pie{360} & \pie{90} & \pie{90} & \pie{90} & \pie{90} & \pie{360} & \pie{360} & \pie{90} & \pie{90} & \pie{90} & \pie{90}\\
AdvDoor~\cite{zhang2021advdoor} & \pie{360} & \pie{90} & \pie{90} & \pie{90} & \pie{360} & \pie{90} & \pie{360} & \pie{90} & \pie{360} & \pie{360} & \pie{90}\\
Bpp~\cite{Wang_2022_bpp} & \pie{360} & \pie{90} & \pie{90} & \pie{90} & \pie{90} & \pie{360} & \pie{360} & \pie{90} & \pie{360} & \pie{90} & \pie{90}\\
LC~\cite{turner2019lc} & \pie{360} & \pie{360} & \pie{90} & \pie{360} & \pie{90} & \pie{90} & \pie{90} & \pie{90} & \pie{90} & \pie{90} & \pie{90}\\
Narcissus~\cite{zeng2023narcisus} & \pie{360} & \pie{360} & \pie{90} & \pie{90} & \pie{360} & \pie{90} & \pie{90} & \pie{90} & \pie{90} & \pie{360} & \pie{90}\\
SSDT~\cite{mo2024ssdt} & \pie{360} & \pie{90} & \pie{360} & \pie{90} & \pie{90} & \pie{360} & \pie{90} & \pie{360} & \pie{90} & \pie{360} & \pie{90}\\
Adap-blend~\cite{qi2022revisitingadaptive} & \pie{360} & \pie{90} & \pie{90} & \pie{90} & \pie{360} & \pie{90} & \pie{90} & \pie{90} & \pie{90} & \pie{360} & \pie{90} \\
DFST~\cite{Cheng2021dfst} & \pie{360} & \pie{90} & \pie{360} & \pie{90} & \pie{360} & \pie{90} & \pie{90} & \pie{90} & \pie{90} & \pie{360} & \pie{90} \\
DFBA~\cite{cao2024dfba} & \pie{90} & \pie{90} & \pie{360} & \pie{360} & \pie{90} & \pie{90} & \pie{90} & \pie{90} & \pie{90} & \pie{360} & \pie{90}\\
\midrule
\ourmethod & \pie{360} & \pie{360} & \pie{360} & \pie{90} & \pie{360} & \pie{90} & \pie{90} & \pie{90} & \pie{360} & \pie{360} & \pie{360}\\
\bottomrule
\end{tabular}
\begin{tablenotes}
    \item \pie{90} the item is not supported by the defense; \pie{360} the item is supported by the defense.
\end{tablenotes}
\end{threeparttable}
\end{table*}

\begin{table*}[htb]
\centering
\caption{Input-space detection results.}
\label{tab:inputlevel}
\begin{tabular}{lcccccccccccc}
\toprule
\multirow{2}{*}{Attack} & \multicolumn{4}{c}{Scale-up~\cite{guo2023scaleup}} & \multicolumn{4}{c}{IBD-PSC~\cite{hou2024ibdpsc}}\\
\cmidrule(lr){2-5}
\cmidrule(lr){6-9}
 ~ & TPR & FPR & AUC & F1 & TPR & FPR & AUC & F1 \\
\midrule
BadNets~\cite{Gu_2019_badnet} & 81.93 & 32.90 & 0.7627 & 0.7524 & 100 & 7.90 & 0.9996 & 0.9606\\
Blend~\cite{chen_2017_blend} & 99.32 & 38.74 & 0.8681 & 0.8275 & 100 & 0.90 & 1.00 & 0.9953\\
Adap-Blend~\cite{qi2022revisitingadaptive} & 68.72 & 18.99 & 0.7621 & 0.7297 & 53.95 & 11.77 & 0.8731 & 0.6495\\
\midrule
\ourmethod (PR=5\%) & 24.40 & 17.69 & 0.5463 & 0.3409 & 0.00 & 10.33 & 0.5698 & 0.0 \\
\ourmethod (PR=1\%) & 18.39 & 17.96 & 0.4879 & 0.2656 & 0.00 & 5.82 & 0.0626 & 0.0\\
\ourmethod (PR=0.5\%) & 7.05 & 16.19 & 0.4034 & 0.1113 & 0.00 & 4.82 & 0.1087 & 0.0\\
\bottomrule
\end{tabular}
\end{table*}

\begin{table*}[htb]
\centering
\caption{\ourmethod against defenses using different architectures on CIFAR10 with a poisoning rate of 5\%. The surrogate indicates the architecture used to generate UPGD as the trigger.}
\label{tab:architectures}
\begin{tabular}{cccccccccccccc}
\toprule
\multirow{2}{*}{Victim} & \multirow{2}{*}{Surrogate} & \multicolumn{2}{c}{No Defense} & \multicolumn{2}{c}{FT-SAM~\cite{Zhu2023ftsam}} & \multicolumn{2}{c}{I-BAU~\cite{zeng2022ibau}} & \multicolumn{2}{c}{FST~\cite{min2023fst}}\\
\cmidrule(lr){3-4}
\cmidrule(lr){5-6}
\cmidrule(lr){7-8}
\cmidrule(lr){9-10}
~ & ~ & BA & ASR & BA & ASR & BA & ASR & BA & ASR \\
\midrule
\multirow{2}{*}{VGG16} & ResNet18 & 92.69 & 95.31 & 92.72 & 78.42 & 90.10 & 14.53 & 89.12 & 92.68\\
~ & VGG16 & 92.57 & 90.10 & 92.22 & 95.14 & 90.20 & 76.51 & 91.72 & 90.58 \\
\midrule
\multirow{2}{*}{DenseNet121} & ResNet18 & 92.39 & 95.62 & 90.98 & 23.88 & 86.73 & 48.14 & 90.77 & 88.94\\
~ & DenseNet121 & 92.38 & 81.07 & 91.10 & 16.91 & 90.90 & 54.76 & 91.13 & 71.29\\
\midrule
\multirow{2}{*}{EfficienNet-B0} & ResNet18 & 87.7 & 96.23 & 84.05 & 71.07 & 87.64 & 95.41 & 82.07 & 97.67\\
~ & EfficienNet-B0 & 86.92 & 92.61 & 83.77 & 71.17 & 86.93 & 92.13 & 82.45 & 68.72\\
\midrule
\multirow{2}{*}{InceptionNeXt-Tiny} & ResNet18 & 85.07 & 91.83 & 85.07 & 2.17 & 85.25 & 91.67 & 82.78 & 3.82\\
~ & InceptionNeXt-Tiny & 85.54 & 96.24 & 85.64 & 90.14 & 85.49 & 97.21 & 83.92 & 97.29\\
\bottomrule
\end{tabular}
\end{table*}

\section{Additional Experiments}

\subsection{Detection of backdoor input}
\label{sec:detect_input}

Backdoor input detection is a defense technique that determines whether or not a given input includes a backdoor trigger.
We show that \ourmethod-generated backdoor samples can resist established backdoor detection methods. 
Table~\ref{tab:inputlevel} shows the input-space detection results using Scale-up~\cite{guo2023scaleup} and IBD-PSC~\cite{hou2024ibdpsc}.
We report the True Positive Rate (TPR), False Positive Rate (FPR), AUC, and F1 score in Table~\ref{tab:inputlevel} for baseline attacks and \ourmethod, where Scale-up and IBD-PSC are effective against three baseline attacks but cannot detect \ourmethod-generated backdoor samples.

\subsection{Different Architectures with Different Surrogate Models}
\label{sec:diff_arch_surrogate}

We evaluate \ourmethod with four additional victim architectures in Table~\ref{tab:architectures}: VGG16, DenseNet121, EfficientNet-B0, and InceptionNeXt-Tiny.
In addition, as \ourmethod requires a surrogate model to generate UPGD as the backdoor trigger, we provide the results when UPGD is generated using different architectures for the surrogate model. 
For each architecture, UPGD is generated by either the victim architecture or ResNet18 to perform our attack.
In Table~\ref{tab:architectures}, we use the three most powerful defenses according to Tables~\ref{tab:pruning} and~\ref{tab:fine_tuning}.
Regardless of the model's architectures or the architectures for UPGD, \ourmethod bypasses most defenses.
This is because the UPGD contains semantic information of the target class and can be transferred among different architectures~\cite{Moosavi2017uap}. 
In a few cases, using UPGD generated by the same architecture shows better attack performance.
For example, conducting \ourmethod on InceptionNeXt-Tiny with UPGD generated by InceptionNeXt-Tiny shows ASRs above 90\%, but also a much lower ASR when using UPGD generated by ResNet18.
We conjecture that transferring UPGD from ResNet18 to InceptionNeXt-Tiny is more difficult than transferring it to other architectures due to the large convolution kernel design of InceptionNeXt.

\subsection{Large and Transformer-Based Architectures}
\label{sec:large_vit}

To further evaluate the effectiveness of \ourmethod on large-sized and especially transformer-based models, we conduct experiments on various architectures, including ResNet101~\cite{He_2016_resnet}, Swin-Base~\cite{Liu2021swin}, Swin-Large~\cite{Liu2021swin}, ViT-Base~\cite{dosovitskiy2021vit}, and ViT-Large~\cite{dosovitskiy2021vit}.
\\
There are two components (UPGD trigger and ABI) of \ourmethod that are necessary to be applied to transformer-based architectures.
The UPGD trigger generation algorithm is compatible with transformer-based architectures, as transformers' gradient information can be used to generate UPGD perturbations.
However, UCLC~\cite{zheng2022clp} for ABI is incompatible with the attention mechanism.
The UCLC is designed for feature channels of convolutional layers, which do not exist in the attention mechanism.
Therefore, we design a tailored novel pruning method against backdoor neurons in the attention-based models.
Specifically, we consider an attention head as a ``neuron'', the new pruning method then calculates the singular value decomposition of each attention head as an index of relevance to the backdoor effect.
The process can also be considered as solving for the approximation of the Lipschitz Constant for each attention head according to~\cite{zheng2022clp}.
Then, the sensitive heads are updated using the following equation:
\begin{equation}
\label{eq:advesarial_injection_head}
    \begin{aligned}
            \vtheta_l^{(k)} := \left\{ 
            \begin{aligned}
                & \vtheta_l^{(k)} \times 0.5, &\sigma(\vtheta_l^{(k)}) > \text{mean}(\sigma(\vtheta_l)) + u \times \text{std}(\sigma(\vtheta_l))\\
                & \vtheta_l^{(k)}, &\text{otherwise,}
            \end{aligned}
            \right.
    \end{aligned}
\end{equation}
where $\vtheta_l^{(k)}$ is the weight of $k_{th}$ head at $l_{th}$ layer, $\sigma$ is the maximum of values in the singular value decomposition matrix of $\vtheta_l^{(k)}$.
We evaluate the performance of \ourmethod on transformer architectures with three fine-tuning-based defenses. 
Table~\ref{tab:largearch} provides the experimental results and demonstrates the effectiveness of \ourmethod on large and transformer architectures.
Notably, the three fine-tuning defenses cannot effectively decrease the ASR of \ourmethod models.
We conjecture this is because injecting a stealthy backdoor into larger models is easier.

The hyperparameters for training ResNet101 are the same as those in Table~\ref{tab:attack_config}.
For transformer models, the number of training epochs is 100 for both CIFAR10 and ImageNet200.
The perturbation budget of UPGD for transformer models is $\epsilon=16$.
The hyperparameters not mentioned are given in Table~\ref{tab:attack_config}.

\begin{table*}[htb]
\centering
\caption{The performance of \ourmethod on large and transformer-based architectures.}
\label{tab:largearch}
\begin{tabular}{cccccccccccccc}
\toprule
\multirow{2}{*}{Dataset} & \multirow{2}{*}{Victim} & \multirow{2}{*}{Surrogate} & \multirow{2}{*}{Parameters (M)} & \multicolumn{2}{c}{No Defense} & \multicolumn{2}{c}{vanilla FT} & \multicolumn{2}{c}{FT-SAM~\cite{Zhu2023ftsam}} & \multicolumn{2}{c}{FST~\cite{min2023fst}}\\
\cmidrule(lr){5-6}
\cmidrule(lr){7-8}
\cmidrule(lr){9-10}
\cmidrule(lr){11-12}
~ & ~ & ~ & ~ & BA & ASR & BA & ASR & BA & ASR & BA & ASR\\
\midrule
\multirow{6}{*}{CIFAR10} & ResNet101 & ResNet18 & 42.51 & 93.75 & 99.22 & 87.37 & 99.92 & 92.78 & 99.04 & 93.65 & 99.77\\
\cmidrule(lr){2-12}
~ & ViT-Base & \multirow{2}{*}{ViT-Base} & 85.27 & 80.24 & 98.80 & 76.43 & 98.72 & 79.50 & 98.10 & 79.77 & 98.73\\
~ & ViT-Large & ~ & 302.60 & 78.92 & 98.89 & 74.47 & 97.53 & 77.98 & 98.56 & 78.53 & 98.53\\
\cmidrule(lr){2-12}
~ & Swin-Base & \multirow{2}{*}{Swin-Base} & 86.70 & 83.08 & 99.65 & 81.61 & 99.80 & 80.21 & 98.88 & 82.44 & 99.86\\
~ & Swin-Large & ~ & 194.93 & 83.04 & 99.69 & 80.79 & 99.35 & 80.99 & 99.47 & 79.31 & 99.53\\
\midrule
\multirow{4}{*}{ImageNet200} & ViT-Base & \multirow{2}{*}{ViT-Base} & 85.95 & 83.41 & 98.36 & 82.08 & 98.86 & 83.48 & 98.24 & 80.99 & 97.84\\
~ & ViT-Large & ~ & 303.51 & 86.08 & 89.46 & 83.19 & 89.07 & 86.06 & 90.52 & 82.36 & 85.35\\
\cmidrule(lr){2-12}
~ & Swin-Base & \multirow{2}{*}{Swin-Base} & 86.95 & 74.68 & 97.82 & 72.56 & 98.29 & 75.75 & 96.11 & 72.29 & 94.77\\
~ & Swin-Large & ~ & 195.30 & 76.99 & 95.36 & 74.29 & 97.91 & 75.35 & 90.77 & 73.07 & 90.72\\
\bottomrule
\end{tabular}
\end{table*}

\subsection{Backdoor analysis by feature space inversion}
\label{appendix:featureloss}

We also provide a feature space analysis for different attacks by using a feature mask to decouple the benign and backdoor features following BTI-DBF~\cite{xu2024btidbf} and BAN~\cite{xu2024ban}.
The decoupling assumes that benign features related to the correct prediction introduce a lower loss, and the backdoor features related to backdoor prediction introduce a higher loss.
In particular, the benign and inversed backdoor features are decoupled as follows:
\begin{equation}
    \label{eq:decoupling}
    \mathop{\min}\limits_{\vm} \sum_{ (\vx,y)\in \Dc_l } \Big[ \mathcal{L} \big( f_L \circ (g(\vx) \odot \vm ), y \big) - \mathcal{L} \big( f_L \circ (g(\vx) \odot (1-\vm), y ) \big) + \lambda | \vm | \Big],
\end{equation}
where $g = f_{L-1} \circ \cdots f_1$ is the model without the classification head.
$\vm$ is the learned feature mask, and $\Dc_l$ is a small set of benign samples with correct labels.
As validated with BTI-DBF~\cite{xu2024btidbf} and BAN~\cite{xu2024ban}, benign and backdoor features can be decoupled by the mask $\vm$, after which backdoor features will introduce a substantially higher loss with respect to the ground truth label.

\begin{figure*}[tb]
\centering
\includegraphics[width = 0.9\linewidth]{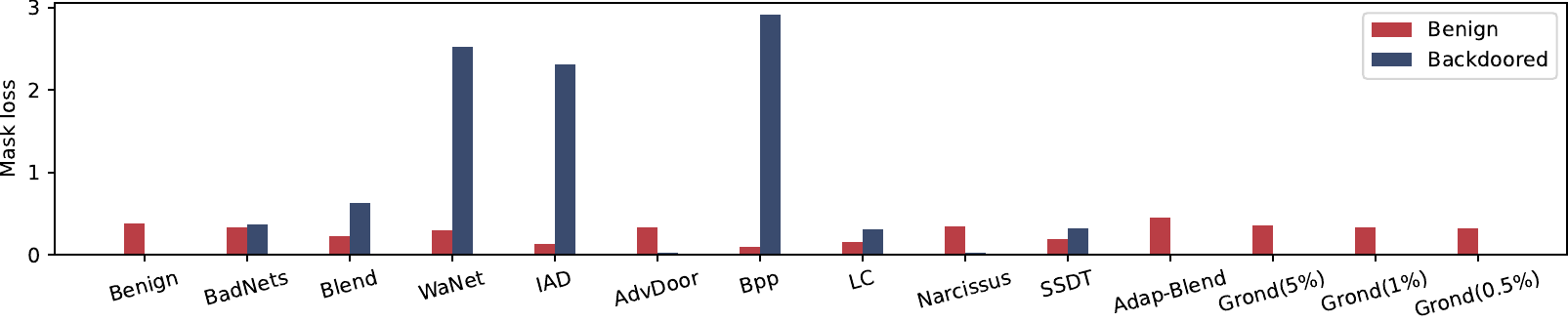}
\caption{Benign and inversed backdoor feature loss (\Eqref{eq:decoupling}) for all baseline attacks. Large backdoored loss indicates that the backdoor is prominent in the feature space.}
\label{fig:featureloss}
\end{figure*}

Decoupled benign feature loss and backdoor feature loss of all evaluated attacks are demonstrated in Figure~\ref{fig:featureloss}, where benign feature loss is represented by the first term in~\Eqref{eq:decoupling} and backdoor feature loss by the second term.
It can be observed that several backdoor attacks introduce prominent backdoor features, such as WaNet, IAD, and Bpp, which result in substantially higher backdoor feature loss than benign feature loss.
In contrast, several backdoor attacks introduce less prominent backdoor features, including AdvDoor, Narcissus, Adap-blend, and \ourmethod.
Revisiting Tables~\ref{tab:pruning},~\ref{tab:fine_tuning}, and~\ref{tab:datasets_allattacks}, attacks that introduce prominent backdoor features are more susceptible to backdoor defenses than attacks with less prominent backdoor features.
Both TAC and feature space inversion analyses further confirm that \ourmethod provides comprehensive stealthiness.


\noindent
\textbf{Hyperparameters for the Inversed Backdoor Feature Loss}
Following the settings in BTI-DBF~\cite{xu2024btidbf} and BAN~\cite{xu2024ban}, 
we use Adam and the learning rate of 0.01 to search for 20 epochs for the feature mask in~\Eqref{eq:decoupling}.
The optimization of the mask uses 1\% of the training data.
$\lambda$ is 0.72.
The elements in the mask are limited to continuous values between 0 and 1.


\begin{table}[tb]
\centering
\caption{Evaluation of dirty-label \ourmethod using ResNet18 on CIFAR10. The dynamic target refers to samples from different classes having different backdoor targets.}
\label{tab:dirtylabel_our}
\begin{tabular}{cccccccccc}
\toprule
\multirow{2}{*}{\tabincell{c}{Dirty-Label \\ \ourmethod}} & \multirow{2}{*}{PR} & \multicolumn{2}{c}{No Defense} & \multicolumn{2}{c}{CLP~\cite{zheng2022clp}} & \multicolumn{2}{c}{FT-SAM~\cite{Zhu2023ftsam}}\\
\cmidrule(lr){3-4}
\cmidrule(lr){5-6}
\cmidrule(lr){7-8}
~ & ~ & BA & ASR & BA & ASR & BA & ASR\\
 \midrule
\multirow{3}{*}{\tabincell{c}{One\\Target}} & 5\% & 91.60 & 100 & 91.41 & 100 & 90.24 & 0.00\\
~ & 1\% & 93.64 & 100 & 91.13 & 40.83 & 91.20 & 46.97\\
~ & 0.5\% & 94.35 & 100 & 91.84 & 97.03 & 91.77 & 99.04\\
\midrule
\multirow{3}{*}{\tabincell{c}{Dynamic\\Target}} & 10\% & 94.26 & 91.95 & 94.13 & 77.29 & 91.67 & 1.46\\
~ & 5\% & 94.22 & 90.43 & 94.04 & 90.34 & 91.45 & 5.48\\
~ & 1\% & 94.33 & 81.82 & 94.26 & 81.80 & 92.36 & 15.77\\
\bottomrule
\end{tabular}
\end{table}

\subsection{Dirty-label \ourmethod}
\label{appendix: dirty_label}
\ourmethod works well in a clean-label setting that uses an invisible trigger and does not change the original labels of the poisoned samples.
However, as we use the supply-chain threat model (the attacker has access to the training process), we could also explore the effect of a dirty-label backdoor attack.
A dirty-label threat model could simplify the backdoor by poisoning samples from any class, while the clean-label is limited to a single class.

Table~\ref{tab:dirtylabel_our} considers two types of dirty-label \ourmethod. 
The ``One Target'' uses the same setting and trigger as clean-label \ourmethod, i.e., misclassifying any inputs with the trigger to the one target class.
The ``Dynamic Target'' refers to misclassifying inputs from different classes to different targets.
In addition, ``Dynamic Target'' \ourmethod uses non-target UPGD as the trigger, which is generated by maximizing the loss between $\vx + \bm{\delta}$ and the true label.

According to the results, One Target \ourmethod achieves very high ASR.
Dynamic Target \ourmethod performs worse than One Target since Dynamic Target is a more complex task.
Overall, dirty-label \ourmethod can still perform well, and even has a higher ASR than the clean-label setting.
However, dirty-label \ourmethod is less robust than clean-label \ourmethod against backdoor defenses because it is less stealthy.
For example, FT-SAM can decrease the ASR of Dynamic Target \ourmethod (PR=5\%) from 94.22\% to below 10\%, and decrease One Target ASR from 100\% to 0.0\%.
We conjecture this is because the dirty-label setting obfuscates the benign semantics of images to their true labels, as the poisoned samples in each class have been assigned the backdoor target label. 
In contrast, only the target class is poisoned in a clean-label setting.

\subsection{Adversarial Backdoor Injection Does Not Impact Backdoor Effectiveness in Case of No Defense}
\label{appendix:advinjection_otherattack}

\begin{figure*}[tbh]
\centering
\includegraphics[width = 1\linewidth]{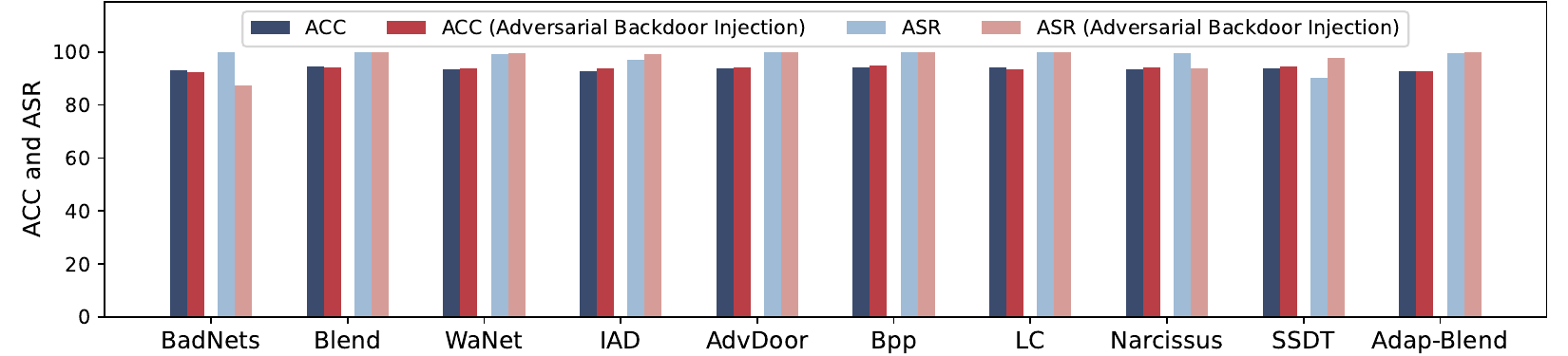}
\caption{The attack performance (no defense) when combined with Adversarial Backdoor Injection.}
\label{fig:clp_otherattacks_nodefense}
\end{figure*}

Figure~\ref{fig:clp_otherattacks_nodefense} shows additional Adversarial Backdoor Injection results against models without defense.
We show that Adversarial Backdoor Injection does not influence the backdoor effectiveness in general when no defense is applied.

\subsection{Further TAC analysis}
\label{appendix:tac_others}

\begin{figure*}[tbh]
\centering
\includegraphics[width = 1\linewidth]{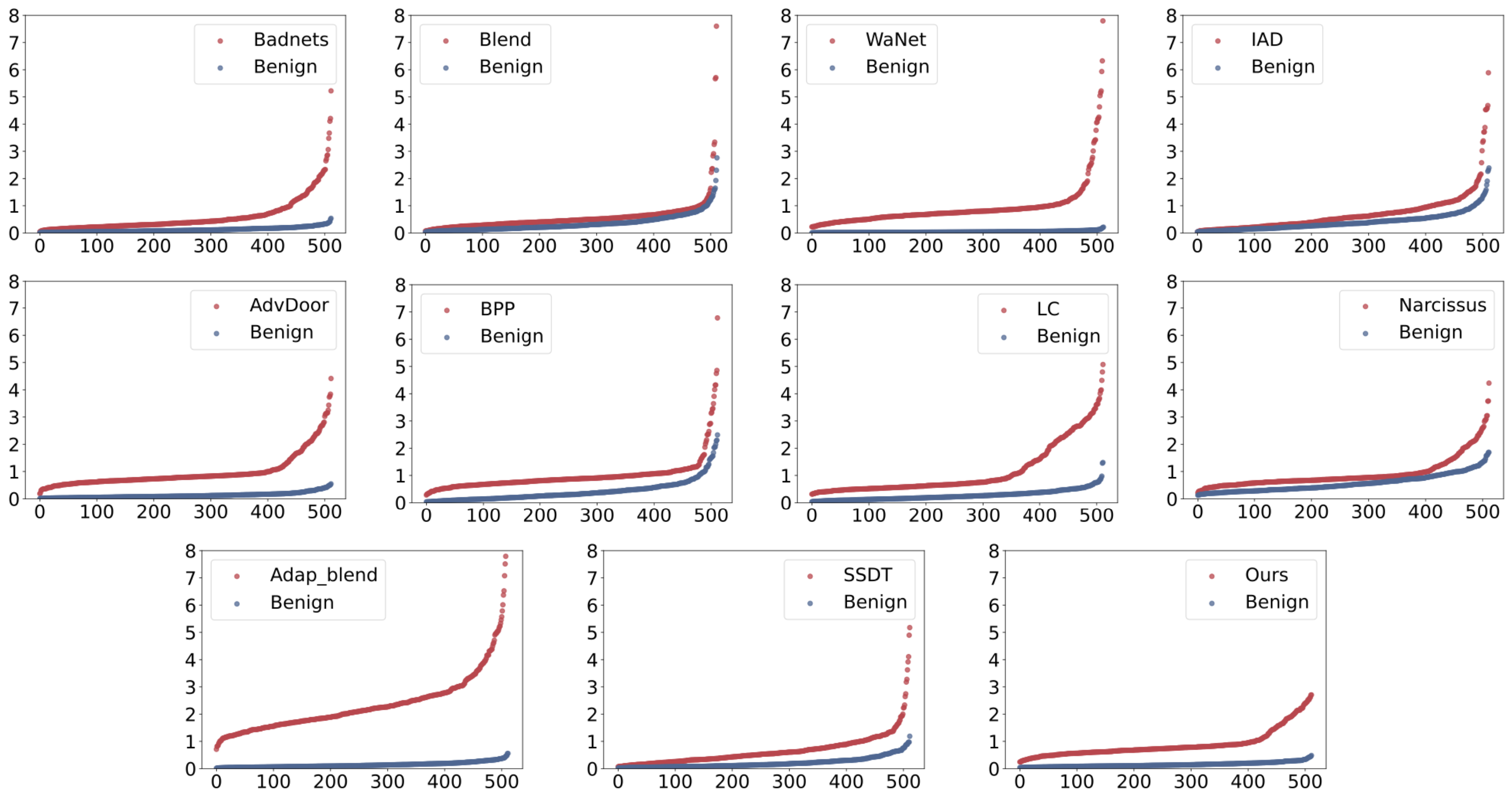}
\caption{TAC plots of sorted TAC values, which show the prominent neurons of baseline attacks. The $y$ axis contains the TAC values, and the $x$ axis is the index of neurons. Prominent neurons are not found in our attack.}
\label{fig:tac_sort}
\end{figure*}

For a clearer demonstration, we also provide sorted TAC value plots in Figure~\ref{fig:tac_sort}, which sort the TAC values in Figure~\ref{fig:tac_all_baseline}.
Figure~\ref{fig:tac_sort} demonstrates the existence of prominent neurons, and \ourmethod is more stealthy.

\begin{table}[tb]
\centering
\caption{Evaluation of \ourmethod against adversarial training (AT) on CIFAR10. RA refers to accuracy under PGD~\cite{madry2018pgd} adversarial attack.}
\label{tab:adversarialtraining}
\begin{tabular}{cccccccccc}
\toprule
\multirow{2}{*}{\tabincell{c}{Arch}} & \multicolumn{3}{c}{No Defense} & \multicolumn{3}{c}{AT}\\
\cmidrule(lr){2-4}
\cmidrule(lr){5-7}
 ~ & BA & ASR & RA & BA & ASR & RA \\
 \midrule
 ResNet18 & 93.43 & 98.04 & 3.82 & 88.56 & 89.69 & 62.24 \\
VGG16 & 92.57 & 90.10 & 9.64 & 85.64 & 85.78 & 59.75\\
DenseNet121 & 92.38 & 81.07 & 9.72 & 86.67 & 64.30 & 58.66\\
\bottomrule
\end{tabular}
\end{table}

\begin{figure*}[tbh]
\centering
\includegraphics[width = 1\linewidth]{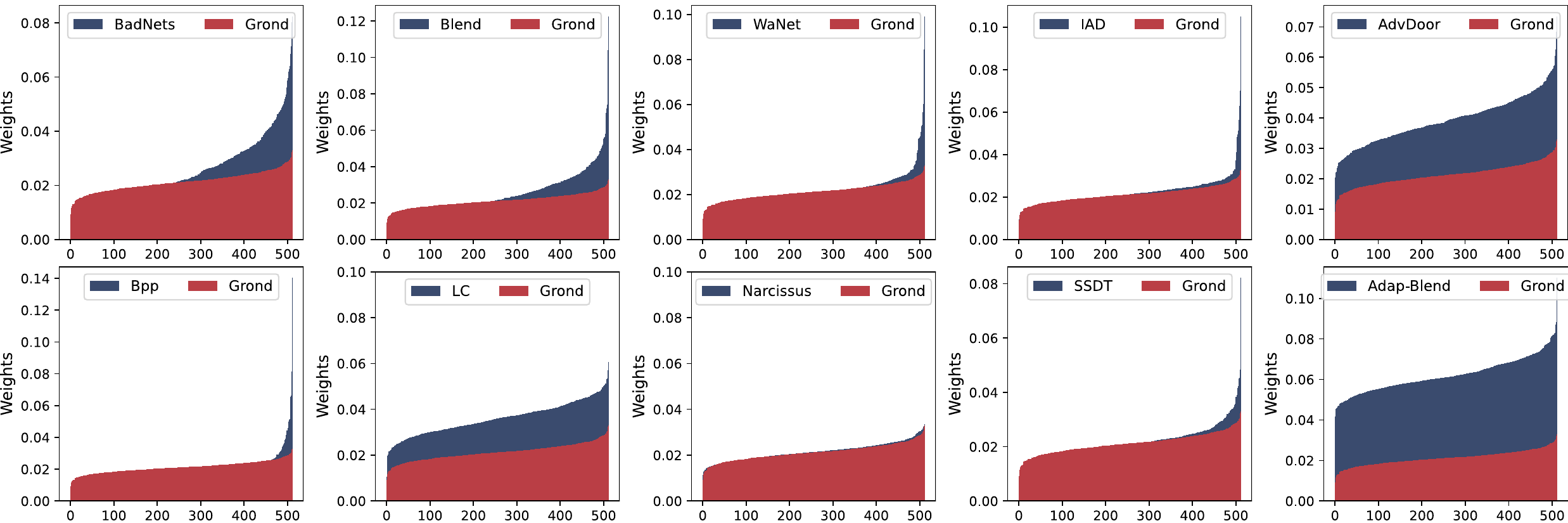}
\caption{The weight changes after FT-SAM~\cite{Zhu2023ftsam} for baseline attacks.}
\label{fig:weight_changes_all}
\end{figure*}

\subsection{Adversarial Training as A Defense}
\label{app:at_defense}
We evaluate \ourmethod with adversarial training (AT), which is considered one of the most successful defenses against adversarial attacks, since the UPGD trigger is a special type of adversarial perturbation.
Table~\ref{tab:adversarialtraining} shows the backdoor performance and robustness of \ourmethod before and after adversarial training using CIFAR10 with three architectures (ResNet18, VGG16, and DenseNet121).
The robust accuracy against PGD attacks significantly increased after AT, confirming AT's effectiveness.
However, AT does not decrease ASR, which means \ourmethod backdoor can resist AT.

\subsection{Weight Changes After FT-SAM with 10 Baseline Attacks}
\label{sec:weight_changes_all}

Figure~\ref{fig:weight_changes_all} provides the results of weight changes after applying FT-SAM~\cite{Zhu2023ftsam} to ten baseline attacks. 
The weight changes in baseline attacks show a significant difference in a few prominent neurons.

\subsection{Attempts to Craft Inputs that Minimize Changes in the Parameter Space}
In this section, we also try to craft poisoned training data that minimizes changes in the parameter space instead of using ABI.

In the previous design of \ourmethod, the backdoor is injected by the UPGD trigger. 
Then, \ourmethod, through ABI, limits the changes of backdoor neurons by reducing their weights.
We consider replacing the ABI component with an operation only in the training data, which does not require control over training.
Our goal is to partially break the effectiveness of injecting the backdoor via the poisoned data.
Specifically, we add poisoned data into the target class with a poisoning rate of $PR=5\%$ under the clean-label setting, following the standard setting in previous work~\cite{wang2023unicorn,xu2024ban}.
Then, we add poisoned data into every class with a poisoning rate of $PR\times\lambda$.
We aim to use the poisoned data in non-targeted classes to partially break the effectiveness of poisoned data in the targeted class.
Table~\ref{tab:input_minimize_parameter} provides the experimental results with different $\lambda$ values.
With the decrease in $\lambda$, both ASR (no defense) and ASR (after applying CLP) increase.
This means that our attempt to minimize changes in the parameter space for stealthiness works.
However, ASR after applying CLP is still lower than 10\%, which means the attack is not robust against the backdoor defense.

\begin{table}[tb]
\centering
\caption{Crafting inputs to minimize changes in the parameter space.}
\label{tab:input_minimize_parameter}
\begin{tabular}{cccccccccc}
\toprule
\multirow{2}{*}{$\lambda$} & \multicolumn{2}{c}{No Defense} & \multicolumn{2}{c}{CLP}\\
\cmidrule(lr){2-3}
\cmidrule(lr){4-5}
 ~ & BA & ASR & BA & ASR \\
 \midrule
 0.5 & 94.09 & 1.43 & 93.39 & 1.43\\
 0.1 & 94.23 & 6.09 & 93.43 & 1.97 \\
 0.01 & 93.69 & 26.83 & 90.90 & 3.87\\
 0.001 & 93.88 & 68.81 & 92.80 & 4.89\\
 0.0005 & 94.13 & 90.78 & 92.81 & 5.84 \\ 
\bottomrule
\end{tabular}
\end{table}

\subsection{Examples of Poisoned Images}
\label{appendix:trigger_images}

\begin{figure}
\centering
\includegraphics[width = 1\linewidth]{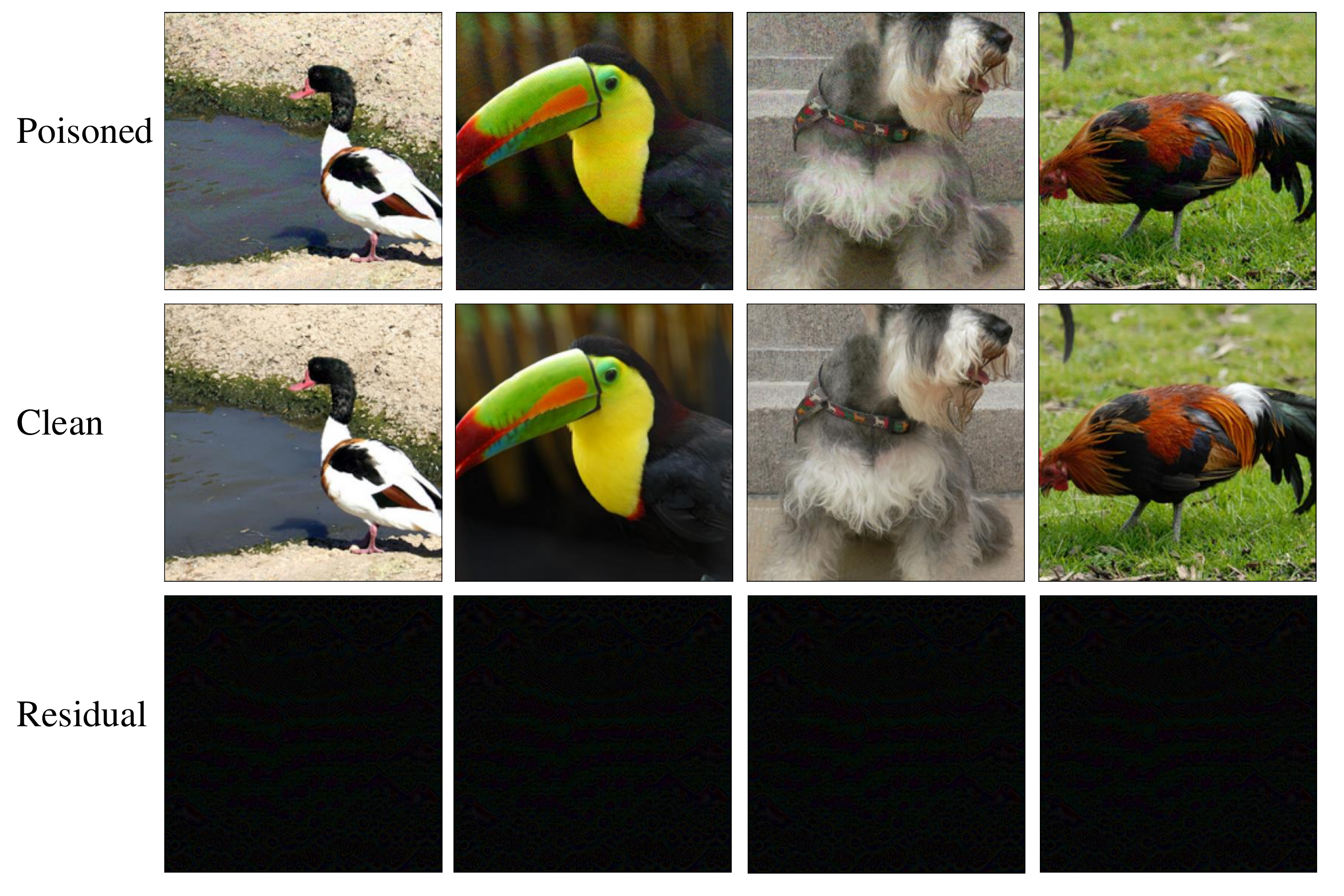}
\caption{Examples of poisoned ImageNet200 images by \ourmethod. We only poison training images from the class ``stingray''  in our experiments with ImageNet200.}
\label{fig:trigger_images}
\end{figure}

Figure~\ref{fig:trigger_images} shows four training images from ImageNet200 when applied with UPGD.
Please note that the images are only meant to demonstrate imperceptible trigger and poisoning perturbations. 
In our experiments, we only poison training images from the class ``stingray'' to inject the backdoor. The first row depicts poisoned images, while the second contains clean ones. Finally, the third row contains the residual of the first two rows. Notice that \ourmethod does not introduce any visible difference to the clean images.

\end{document}